\documentclass[sigconf]{acmart}

\usepackage{graphicx} % Required for inserting images
\usepackage{tabularx}
\usepackage{subcaption}

\copyrightyear{2025}
\acmYear{2025}
\setcopyright{cc}
\setcctype{by}
\acmConference[ARTECH 2025]{12th International Conference on Digital and Interactive Arts : ARTECH 2025 – Media Art Cultures, Communities \& Territories}{November 26--28, 2025}{Braga, Portugal}
\acmBooktitle{12th International Conference on Digital and Interactive Arts : ARTECH 2025 – Media Art Cultures, Communities \& Territories (ARTECH 2025), November 26--28, 2025, Braga, Portugal}
\acmPrice{}
\acmDOI{10.1145/3773699.3773918}
\acmISBN{979-8-4007-2001-7/25/11}

\title{Protocol as Poetry: A Case Study of Pak's Smart Contract-Based Protocol Art}

\author{Botao Amber Hu}
\orcid{0000-0002-4504-0941}
\affiliation{%
  \institution{University of Oxford}
  \city{Oxford}
  \country{UK}
}
\email{botao.hu@cs.ox.ac.uk}

\begin{abstract}
Protocol art has recently proliferated through blockchain-based smart contracts, building on a century-long lineage of conceptual, participatory, interactive, systematic, algorithmic, and generative art practices. Few studies have examined the characteristics and appreciation of this emerging art form. To address this gap, this paper presents an annotated portfolio analysis of protocol artworks by Pak, a pioneering and influential pseudonymous artist who treats smart contracts as medium and collective participation through protocol as message. Tracing the evolution from early open-edition releases of \emph{The Fungible} (2021) and the dynamic mechanics of \emph{Merge} (2021) to the soul-bound messaging of \emph{Censored} (2022) and the reflective absence of \emph{Not Found} (2023), we examine how Pak choreographs distributed agency across collectors and autonomous code, demonstrating how programmable protocols become a social fabric in artistic meaning-making. Through thematic analysis of Pak's works, we identify seven core characteristics distinguishing protocol art from other art forms: (1) system-centric rather than object-centric composition, (2) autonomous governance enabling open-ended control, (3) distributed agency and communal authorship, (4) temporal dynamism and lifecycle aesthetics, (5) economy-driven engagement, (6) poetic message embedded in interaction rituals, and (7) interoperability enabling composability for emergent complexity. We then discuss how these features set protocol art apart from adjacent movements such as conceptual, generative, participatory, interactive, and performance art. By analyzing principles grounded in Pak's practice, we contribute to the emerging literature on protocol art (or ``protocolism'') and offer design implications for future artists exploring this evolving form.

\end{abstract}

\begin{teaserfigure}
    \centering
        \includegraphics[width=1\linewidth]{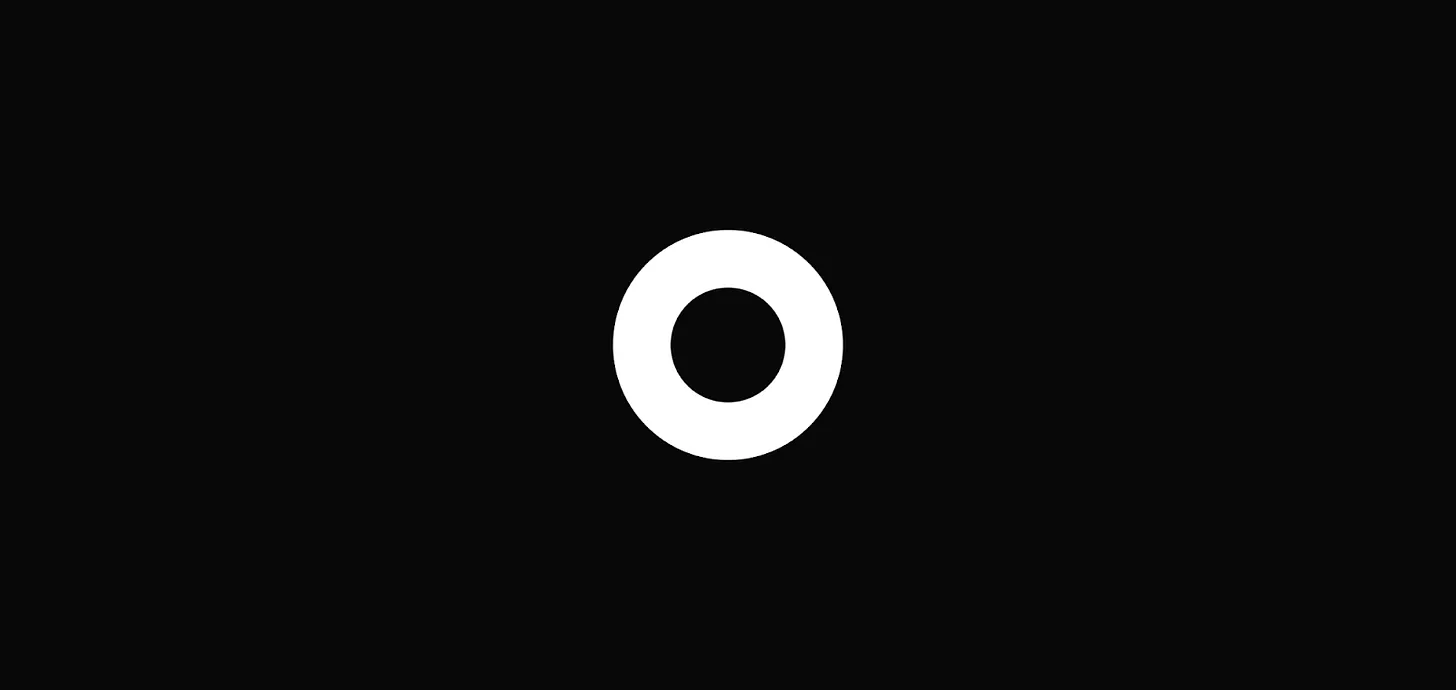}
        \Description{A white torus (ring shape) centered on a solid black background, serving as the recurring visual identity of pseudonymous digital artist Pak.}
        \caption{Pak's signature symbol, a minimalist white ring on black known as ``The Nothing,'' used as the artist's persistent identity across social media and NFT platforms.}
    \label{fig:pak}
\end{teaserfigure}



\begin{document}

\maketitle

\section{Introduction}

Over the last decade, the advent of blockchain networks, smart contracts, and non-fungible tokens (NFTs) has transformed digital culture by relocating trust from central institutions to cryptographic code---famously encapsulated in the motto \textit{``code is law''} \cite{lessig2000code}. Beyond reshaping finance through decentralized finance (DeFi), these technologies have begun to reconfigure artistic practice. An emerging cohort of artists now treat smart contracts as their medium, designing works whose meaning unfolds through on-chain participation. Pioneering projects such as Paul Seidler's \textit{terra0} (2016)\footnote{\url{https://terra0.org/}} \cite{terra02016}, which frames a forest as a self-owning economic entity \cite{seidler2017terra0}, Sarah Friend's socially-driven on-chain \textit{Lifeforms} (2021)\footnote{\url{https://lifeforms.supply/}} \cite{lifeforms2021}, and Primavera De Filippi's robo-botanical \textit{Plantoid} series (2015)\footnote{\url{https://plantoid.org/}} \cite{plantoid2017} illustrate how smart-contract logic can serve as a medium to distribute creative agency through participation. In these works, \emph{``the aura of the artist is on maintaining the process''} \cite{Filippi2025Protocolism}, as code and crowd together generate the artistic outcome. Recent exhibitions such as \textit{World Computer Sculpture Garden} (2024)\footnote{\url{https://worldcomputersculpture.garden/}} \cite{wcsg2024} have further foregrounded this shift, showcasing smart contracts as sculptural pieces that exist immutably on Ethereum: \textit{``This permissionless computing platform has been reimagined as one, functioning simultaneously as archive, gallery, and economy.''} \cite{maltefr2024Computation} Together, these examples signal the emergence of what is increasingly termed \textit{protocol art}.

Yet the concept of protocol art remains theoretically unsettled. De Filippi \cite{Filippi2025Protocolism} defines protocol art as an art form where \textit{``protocols serve as the bridge between two pivotal figures: the original artist (i.e., the `protocol artist') and a diverse array of executors (both human and non-human).''} These protocols \textit{``comprise sets of instructions and constraints meticulously crafted to guide the production of unique artworks''} \cite{Filippi2025Protocolism}. While evocative, this definition is very broad in practice. It is difficult to distinguish protocol art from other long-standing rule-based art forms. For example, instruction-based conceptual art---such as Sol LeWitt's \cite{lewitt1967paragraphs} procedural \textit{Wall Drawing} series \cite{LeWitt2015Sol}, Yoko Ono's participatory \textit{Cut Piece} (1964) \cite{Ono1964Cut}, or Marina Abramović's durational \textit{The Artist Is Present} (2010) \cite{Abramovic2010Artist}---relies on rules and audience action. Generative software artworks follow similar principles, including Vera Molnár's early algorithmic art and contemporary AI-generated pieces \cite{Molnar1975Aesthetic}. Recent blockchain-based generative NFT platforms such as Art Blocks\footnote{\url{https://artblocks.io/}} invite collectors to mint their own algorithmic editions---originally generated from art code in smart contracts. Although some pieces are interactive, allowing interaction through webpages, they rarely involve dynamic interaction and evolution at the smart-contract level once minted. \textit{``The blockchain is not a distribution channel for images but rather the material of the art itself.''} as curator Malte Rauch (@maltefr) argued in his introductory essay to \textit{World Computer Sculpture Garden} \cite{maltefr2024Computation}. A gap persists between broad definitions of protocol art and the specific protocol-art practices now emerging from the use of smart contracts as artistic medium.

Our research question is: \emph{What distinct characteristics define protocol art as evidenced through real-world practices?} 

We address this gap through an annotated portfolio analysis \cite{Gaver2012Annotated, Culen2020Strategies} of the pseudonymous artist \textbf{Pak}\footnote{\url{https://x.com/muratpak}}, who is widely credited with establishing protocol art's foundational vocabulary. Our focus on a single artist is methodologically purposive rather than a sampling limitation: in emerging fields, paradigmatic cases---exemplars that establish the conventions and possibilities of a new form---merit intensive single-case analysis. Just as Duchamp's readymades warranted deep study to understand conceptual art's foundations, or as John Cage's compositions required close analysis to theorize indeterminacy in music, Pak's protocol artworks represent the paradigmatic instantiation of smart-contract-as-medium practice. Pak did not merely participate in protocol art---Pak effectively \emph{invented} its vocabulary: open-edition time-scarcity (\emph{X}, 2020), deflationary merging mechanics (\emph{Merge}, 2021), burn-to-mint economies (\emph{burn.art}, 2021), soul-bound protest tokens (\emph{Censored}, 2022) \cite{Ohlhaver2022Decentralized}, and tokenized absence (\emph{Not Found}, 2023). These mechanisms have since been widely imitated across thousands of NFT projects but never surpassed in conceptual sophistication. Each work demonstrates how participatory on-chain interaction can deliver poetic meaning: collectors are not merely purchasers but co-authors who experience the artwork through its evolving mechanics---whether through token fusion, auto-merging mass, or hidden messages revealed only under certain conditions. This participatory design transforms algorithmic logic into a collective movement that prompts social reflection and generates significant cultural and economic impact. Notably, \emph{Merge} realized \$91.8 million in primary sales in December 2021, heralded at the time as \textit{``the largest-ever public sale of an artwork by a living artist''} \cite{BlockPAKs, Thomas2021Pak}. By embedding metaphors directly in code while enlisting thousands of participants, Pak's oeuvre offers a fertile---and uniquely well-documented---corpus for examining the intertwined dynamics of protocol mechanics, collective engagement, and conceptual intent.

In the remainder of this paper, we analyze Pak's key works to derive common characteristics of protocol art. We present seven salient characteristics that recur across Pak's protocol artworks, which together form what we term Pak's \textit{protocol aesthetic}. Then, we discuss how these traits distinguish protocol art from adjacent art movements such as conceptual, systemic, generative, participatory, interactive, and performance art, and we outline limitations of our single-artist study along with directions for future research. We conclude by reflecting on how protocol art (or ``protocolism'') expands the discourse of digital art and offers new opportunities for artists to engage code, community, and concept as one autonomous system.

\section{Background}

\subsection{Emerging Protocol Arts}
Web3 technologies---blockchains, smart contracts, and non-fungible tokens (NFTs)---reconfigure the terms of authorship by shifting enforcement from institutions to code \cite{Sidorova2025NFTs}. Smart contracts are self-verifying, self-executing, tamper-resistant programs that guarantee rule compliance without a central authority \cite{Wang2018Overview}. NFTs assign cryptographic provenance and ownership to digital artifacts. Together, these technologies allow artists to publish works that run autonomously and invite permissionless participation. As collectors interact on-chain---minting, trading, or triggering contract functions---the artwork's state can evolve in real time, making the protocol, rather than any single static output, the true locus of creativity \cite{Wu2025Rethinking, Guan2025From}. Smart-contract infrastructure enables this participatory evolution at global scale. Chris Dixon argues in \textit{``Read Write Own''} \cite{Dixon2024Read} that ownership distinguishes Web3 from Web2's write and Web1's read---blockchain networks grant power and economic benefits to communities of users. By reshaping ownership, web3 has revived and radically extended the lineage of participatory art \cite{bishop2023artificial}: art in which the public co-owns and co-produces the piece through autonomous participation governed by protocols.

Several early projects foreshadowed the protocol art paradigm. \textit{terra0} (2016) presented a vision of an artwork that manages itself: a forest equipped with sensors and coded as a Decentralized Autonomous Organization (DAO) that sells its lumber rights to fund its own growth \cite{terra02016, Seidler2025Artwork}. Primavera De Filippi's Plantoid (first created in 2015) \cite{plantoid2017} embeds a robotic ``flower'' sculpture in Ethereum such that donors collectively ``pollinate'' the piece with cryptocurrency to fund the creation of new offspring sculptures. By 2024, exhibitions such as \textit{World Computer Sculpture Garden} \cite{wcsg2024} showcased artists (among them 0xhaiku, Absent, Loucas Braconnier (Figure31), Sarah Friend, Material Protocol Arts, Rhea Myers et al., Paul Seidler) whose works exist solely as on-chain contracts, reinforcing the idea that the protocol itself can be the artistic material \cite{maltefr2024Computation}. Rhea Myers's collected writings further document how smart contracts have been used as an artistic medium since 2011 \cite{Myers2023Proof}. 

The pseudonymous artist \textbf{Pak}---also known as Murat Pak---is best known for harnessing blockchain smart contracts as an artistic medium. Early work included creating the AI-curation bot Archillect\footnote{\url{https://archillect.com/}} in 2014. Pak shifted to smart contracts in 2020 and quickly became one of the field's most commercially successful and conceptually adventurous figures. The artist's anonymity (some speculate ``Pak'' may be a collective) and code-driven practice challenge traditional ideas of authorship and ownership. Record-breaking projects such as Merge have placed Pak among the highest-grossing living creators in any medium \cite{Thomas2021Pak}. Pak's projects innovate in protocol mechanisms, inviting thousands of collectors to fuse tokens, burn assets for new tokens, redact and reveal messages, or accumulate ``mass''---demonstrating that participatory creativity can be both conceptually rich and commercially significant.

Collectively, these examples point to an emerging genre---protocol art---in which the rules encoded on-chain constitute the work, and the audience's transactions constitute its performance.

\subsection{Defining Protocol Art?}
Multiple disciplines offer definitions of protocol. In computer networking, a protocol is \textit{``a set of rules for formatting and processing data,''} a common language that enables coordination among machines \cite{Tanenbaum1981}. Media theorist Alex Galloway famously broadened this concept: ``protocol'' is the principle of organization that governs distributed systems---\textit{``how control exists after decentralization''} \cite{galloway2001protocol}. Venkatesh Rao and colleagues, in the Summer of Protocols\footnote{\url{https://summerofprotocols.com/}} research initiative, describe protocols as \textit{``a stratum of codified behavior from which complex coordination can emerge''} \cite{Rao2024Unreasonable}. The material embodiment of a protocol can range from written rules to cryptographically enforced software; what varies is the \textit{``hardness of trust''}---i.e. how difficult it is to deviate from the rules \cite{Stark2023Atoms}.

Within the arts, in her essay \textit{``Protocolism: The Evolving Landscape of Art in the Age of AI''} \cite{Filippi2025Protocolism}, originally published in \textit{All Media is Training Data}, protocolism has been advanced by De Filippi to describe practices that use protocols to choreograph diverse human and non-human executors in the production of artworks \cite{Filippi2025Protocolism}. However, as noted above, this umbrella definition can be too broad when applied to art. Are instruction-based conceptual works like Sol LeWitt's wall drawings ``protocol art''? One might argue yes: LeWitt often did not execute his own wall drawings---he sold the ``recipe'' or instructions (a kind of artistic protocol) to be faithfully executed by others, meaning the artwork is essentially a set of rules carried out by participants. Similarly, is a classic participatory performance piece like Marina Abramović's \textit{The Artist Is Present} (2010) \cite{Abramovic2010Artist} a protocol artwork? The piece had a simple rule: viewers took turns sitting silently across from the artist for a duration, exploring presence and connection. One could view this as a human protocol enacted in a gallery. Even modern generative NFT platforms such as Art Blocks, where collectors mint outputs from pre-written algorithms, involve protocols of creation. What, then, distinguishes protocol art in the context of Web3?

Following the essay \textit{``Computation in the Expanded Field''} \cite{maltefr2024Computation}, Malte Rauch argues that true protocol artists \textit{``are primarily concerned with foundational structures or sets of rules that precede and define a computing environment''}. In these works, \textit{``The blockchain is not a distribution channel for images but rather the material of the art itself.''} The artistic focus shifts from static outputs to the system of rules \cite{Rivero2024Art, Poposki2024Crypto}. In other words, what sets protocol art apart is that the rule-set autonomously generates and governs the art experience over time, without further intervention. Traditional conceptual or generative art \cite{boden2009generative} uses instructions or code to produce a finished piece---perhaps an interactive one. Protocol art, however, uses instructions or code to create an open-ended process or system that invites collective participation through co-creation of meaning and co-ownership of the piece. As we will show through Pak's work, protocol art emphasizes system-centric creation, persistent autonomy, communal participation, temporal evolution, integrated economic and social incentives, and cross-project interoperability---together resulting in a distinctive aesthetic and conceptual stance. However, the terrain of protocol art continues to be shaped as more artists join this movement, with emerging critical \cite{Poposki2025Critical, Poposki2024Corpus} and ethical \cite{Calvo2024Cryptoart} perspectives enriching the discourse. Defining protocol art is still far from consensus and remains an ongoing process.

\section{Method}

To investigate the emerging characteristics of protocol art, we conducted an annotated portfolio analysis \cite{Gaver2012Annotated} of Pak's nine selected landmark protocol artworks, listed in Table \ref{tab:pak-projects}. We compiled data from four complementary strata: (1) On-chain evidence---we examined smart-contract source code and transaction logs for every major Pak project (using resources like Etherscan\footnote{\url{https://etherscan.io/}} and Dune Analytics\footnote{\url{https://dune.com/}}); (2) Project interfaces---we reviewed official launch webpages, marketplace listings (e.g., Nifty Gateway\footnote{\url{https://niftygateway.com/}} drop pages), and any interactive frontend elements; (3) Discourse layer---we gathered journalistic articles, critical essays, and high-signal social media threads (e.g., Twitter analyses) that captured community reception and interpretation; (4) Authorial voice---we collected Pak's own statements from archived interviews and tweets to understand the artist's intent in each project.

\paragraph{Corpus Selection.} We assembled a portfolio of Pak's protocol artworks using purposive sampling with explicit inclusion criteria: (1) the work's primary medium is smart contract mechanics, not merely NFT-based distribution of pre-existing digital images; (2) the work involved documented collective participation, operationalized as more than 100 unique wallet addresses interacting with the contract; (3) the work generated substantial critical discourse, operationalized as at least three independent published analyses in art journalism, academic venues, or high-engagement social media threads. Nine works met all criteria (Table~\ref{tab:pak-projects}). We excluded Archillect (Pak's 2014 AI curation bot, which predates blockchain and lacks protocol art characteristics) and minor releases without significant documentation or collective participation.

\paragraph{LLM-Assisted Data Gathering.} We broadened coverage by deploying a large-language-model agent (ChatGPT-o3, March 2024 version) to aggregate publicly available references to each artwork. The prompt template was: ``Compile all publicly documented information about [artwork name] by Pak, including: official announcements, smart contract mechanics, critical reception, and participation statistics. Cite sources.'' LLM outputs served as initial summaries requiring human verification. The first author independently verified all factual claims against primary sources: smart contract code on Etherscan, official project websites, archived tweets, and transaction records on Dune Analytics. Discrepancies were resolved in favor of primary sources.

\paragraph{Thematic Analysis.} Guided by theoretical sampling, we subjected the aggregated dataset to iterative open coding. Initial codes were generated inductively from the first three works (\emph{X}, \emph{The Title}, \emph{The Fungible}), yielding 23 preliminary codes (e.g., ``time-based scarcity,'' ``collector-as-author,'' ``economic gamification''). These codes were refined through constant comparison with subsequent works, guided by prior scholarship on conceptual art \cite{lewitt1967paragraphs}, systems art \cite{burnham1968systems}, generative art \cite{boden2009generative}, participatory art \cite{bishop2023artificial}, and interactive art \cite{seevinck2017emergence}. After analyzing \emph{Merge} (the fourth work), codes stabilized into seven analytical categories: (1) Concept; (2) System--mechanism dynamics; (3) Participatory interaction; (4) Artist's control; (5) Collective emergent behavior; (6) Poetic meaning-making; (7) Key statistics; and (8) Annotation. The LLM agent assisted in initial classification under the first seven lenses (see Appendix~A for compiled entries), after which one human researcher manually verified, corrected, and enriched every entry with additional details from primary sources. The eighth lens, Annotation, was composed entirely by the human researcher, mapping each work to the seven characteristics identified in Section~\ref{sec:results} and providing interpretive commentary grounded in the verified data.

During focused coding, we refined the properties and dimensions within each lens and constructed a comparative matrix of all projects (distilled in Table \ref{tab:pak-projects}). We then performed axial coding to trace connections across lenses and identify higher-order themes and consistent patterns. This process yielded a grounded characteristic model of Pak's protocol art practice that structures the discussion that follows.

\begin{table*}[htbp]
\centering
\scriptsize
{
\setlength{\tabcolsep}{3pt}
\setlength{\extrarowheight}{2pt}
\begin{tabularx}{\textwidth}{X|X|X|X|X|X|X}
\toprule
\textbf{Project (Year)} &
\textbf{Description} &
\textbf{Participatory Interaction} &
\textbf{System Mechanism} &
\textbf{Artist's Control} &
\textbf{Participants / Scale} &
\textbf{Collective Emergent Behavior} \\
\midrule

X (2020) &
Introduced time-based scarcity with a 24-hour unlimited open edition, shifting value from object to moment. &
24-hour open-edition mint; collectors could mint unlimited copies within the time window. &
Time-based scarcity: supply fixed by duration, not by a preset count. &
None after launch (rules immutable once deployed). &
15 NFT pieces; 13 open editions (61 editions minted) + 2 one-of-one auctions (unique pieces). Hundreds of wallets participated. &
``FOMO''-driven rush as deadline approached; final supply decided by community minting behavior. \\\hline

The Title (2021) &
Identical image sold under differently priced on-chain names, showing metadata (title/pricing) can dictate value more than pixels. &
Fixed-price and auction sales of visually identical NFTs, each bearing a different on-chain title and price; one piece intentionally left unsold. &
Value experiment: image constant, metadata (title + scarcity) varies per token. &
Only predetermined prices/titles; tokens immutable after mint. &
9 token types (6 sold via various formats, 1 remained unsold, plus gifts); \textasciitilde300 total editions; dozens of collectors across formats. &
Community debate on what confers an NFT's value (image vs.\ narrative vs.\ price); social media buzz about who obtained ``\emph{The Expensive},'' etc. \\\hline

The Fungible (2021) &
Sotheby's open-edition turned fungibility into spectacle: thousands of interchangeable cubes questioned uniqueness. &
Hybrid drop: open-edition ``Cube'' tokens minted in large quantities + puzzle/leaderboard rewards during a Sotheby's event. &
Fungible cubes could be aggregated into higher-tier NFTs; gamified rewards for top accumulators and puzzle solvers. &
Rules and reward tiers fixed in contract; artist manually validated puzzle answers and distributed bonus NFTs. &
23{,}598 cubes bought by 3{,}080 wallets $\rightarrow$ yielded 6{,}165 unique NFTs (after merging by quantity). &
Competitive accumulation (to get larger cube clusters); cooperative puzzle-solving; viral social media tasks as part of the game. \\\hline

burn.art (2021) / \$ASH &
Perpetual ``creation-through-destruction'': burn any NFT to mint the social currency \$ASH, gamifying scarcity. &
Ongoing: anyone can burn any NFT on the platform and receive \$ASH (ERC-20 token) via a bonding-curve formula. &
``Creation via destruction'': burn an NFT, get fungible \$ASH; token yield halves at supply milestones. &
Contract rules fixed; artist later released drops purchasable only with \$ASH, influencing demand. &
Thousands of NFTs burned; thousands of \$ASH holders emerged. &
Emergent burn economy: holders strategized which NFTs to sacrifice; a subculture of ``burners'' formed, embracing destruction as creation. \\\hline

Lost Poets (2021--22) &
Multi-stage strategy game of 65{,}536 AI-generated Pages evolving or burning into Poets; collectors co-wrote an ever-shifting narrative. &
Multi-stage narrative game: 65{,}536 ``Page'' NFTs sold; holders could burn Pages for \$ASH or transform Pages into AI-generated ``Poet'' NFTs; on-chain naming of Poets. &
Multi-act smart contract: burn-vs-mint branching, periodic airdrops of 1{,}024 special ``Origin'' Poets to top participants; progressive trait reveals over time. &
Pak triggered each Act (I--IV) via contract calls and airdrops, but could not reverse user choices or outcomes once each phase started. &
Approx.\ 8{,}000 unique holders; resulted in approx.\ 28{,}000 Poet NFTs (others remained as Pages or were burned). &
Rich strategy debates (whether to keep Pages, burn for \$ASH, or transform to Poets); ``naming guilds'' formed to coordinate thematic naming; puzzle-hunting for hidden clues in traits. \\\hline

Invisible Mechanism (a.k.a.\ Hate) (2021) &
Premiered a \texttt{move()} contract granting unilateral token relocation by the artist---an audacious take on creator sovereignty. &
Airdrop performance: 30 of Pak's outspoken critics were invited to provide wallet addresses and received non-transferable ``Hate'' NFTs. &
Custom Move contract: tokens soul-bound to recipient (non-transferable), with an admin-only \texttt{move()} function letting the artist seize or relocate tokens at will. &
Absolute control: artist could reclaim or change tokens at any time; recipients had no control. &
30 forced participants (recipients of the airdrop). &
Sparked wide discussion on ownership and power asymmetry in NFTs; recipients tried and failed to remove tokens; community debated the ethics of an all-powerful contract. \\\hline

Merge (2021) &
312{,}686 ``mass units'' sold that algorithmically fused into larger single tokens after transfers---ownership as live sculpture. &
48-hour open sale of ``mass units''; every buyer received a single NFT whose ``mass'' automatically increased as they bought or acquired more units. &
Self-merging deflationary tokens: all units in one wallet fused into one token; built-in leaderboard tiers (numbered 1--4 by color, with a dynamic ``Alpha'' class for the top holder). &
Rules fixed in code; artist only involved post-sale via orchestrating an ARG and acknowledging top holders. &
28{,}983 buyers; 312{,}686 total mass units sold $\rightarrow$ resulted in 28{,}983 NFTs at sale end (one per buyer). &
Frenzied ``mass hoarding'' competition; after sale, mergers between wallets reduced token count (``extinction game''); a DAO (``Internship'') even formed to collectively purchase the largest mass (Alpha). \\\hline

Censored (2022) &
With Julian Assange: DAO-funded 1/1 ``Clock'' counting prison days plus open-edition Messages---provenance weaponized for protest. &
Two-part project: (a) a one-of-one Clock auction via DAO, (b) a 48-hour open-edition Message mint where anyone could submit any text at any price. &
(a) Clock is a dynamic counter tied to Assange's status (days in prison, now days free since his June 2024 release); (b) Message NFTs initially showed each text as blacked-out and were revealed upon Assange's release. &
Clock runs autonomously; reveal was bound to Assange's release (triggered June 2024); messages are immutable once written (Pak cannot alter or censor them). &
Clock was won by AssangeDAO (10{,}000+ members collectively own it); 29{,}766 Message NFTs were minted (one per wallet). &
Grass-roots mobilization: a global DAO formed to bid for Clock; thousands minted censored messages creating a ``censored chorus.'' Shared anticipation of the potential reveal unites participants; project bridged crypto art and political activism. \\\hline

Not Found (\#404) (2023) &
A 1-of-1 NFT deliberately devoid of metadata (token ID 404), transforming ``page not found'' into a meditation on digital absence. &
Single 1-of-1 charity auction; NFT token has no image or metadata (only token ID ``404'' visible). &
``Deliberate absence'': token's metadata contract returns the string ``In memory of the absent'' instead of standard metadata, referencing the HTTP 404 error via its token ID. &
Designed absence: artist coded the contract to return only a memorial string in place of metadata; the metadata contract is technically mutable but left empty by intent. &
1 owner (noted collector WhaleShark won the auction). &
Community reflection on owning ``nothing''; sparked discourse on whether a void can be art. Served as a memorial tribute (for the late Alotta Money) that many in the community shared and discussed. \\
\bottomrule
\end{tabularx}
}
\caption{Comparative overview of selected Pak projects}
\label{tab:pak-projects}
\end{table*}

\section{Results: Characterizing Pak's Protocol Arts}
\label{sec:results}
Our analysis reveals that Pak's work forms a coherent yet evolving family of protocol artworks. Their aesthetic power lies not in visual form but in the programmable conditions that orchestrate collective action. Across nine landmark projects---\textit{X} (2020)\footnote{\url{https://opensea.io/collection/x-by-pak}}, \textit{The Title} (2021)\footnote{\url{https://niftygateway.com/collections/thetitle}}, \textit{The Fungible} (2021)\footnote{\url{https://www.sothebys.com/en/digital-catalogues/the-fungible-collection-by-pak}}, \textit{burn.art} (2021)\footnote{\url{https://burn.art}}, \textit{Lost Poets} (2021)\footnote{\url{https://lostpoets.xyz/}}, \textit{Invisible Mechanism} (2021), \textit{Merge} (2021)\footnote{\url{https://www.niftygateway.com/collections/pakmerge/}}, \textit{Censored} (2022)\footnote{\url{https://opensea.io/collection/censored-pak-assange}}, and \textit{Not Found} (2023)\footnote{\url{https://opensea.io/collection/in-memory-of-the-absent}}, we observe a noticeable set of characteristics. Together, these characteristics constitute what we call Pak's \textit{protocol aesthetic}---a design grammar that transforms blockchain infrastructure into a stage for large-scale participatory meaning-making. We articulate this aesthetic through seven interlinked characteristics, elaborated in the subsections below.

\subsection{System-centric rather than object-centric composition}
Pak's practice exemplifies a shift from creating discrete art objects to designing dynamic systems. In a \emph{``systems aesthetic''} as theorized by Jack Burnham in the late 1960s \cite{burnham1968systems}, the artist defines goals, rules, and interactions rather than crafting a fixed artifact. Unlike a static painting or sculpture with defined forms, a system-based artwork can evolve over time and respond to external inputs. Pak explicitly embraces this ethos. Pak has argued in interviews and on social media that an NFT's image is merely one element of a token's metadata, and that focusing only on the visual aspect is ``\textit{a naïve frame of view}'' (implying that the structure and behavior encoded in the token are equally part of the art). The true medium for Pak is the underlying protocol---``\textit{the structure of the NFT market itself}''  \cite{Droitcour2021Betting}, which Pak leverages as a creative material \cite{Poposki2024Crypto}.

This system-centric approach is evident in works like \emph{The Fungible} (2021). Rather than a singular digital image, \emph{The Fungible} was presented as a three-day performative market game on Nifty Gateway. Buyers effectively co-created the outcome: they could purchase unlimited open-edition ``Cube'' tokens, and \emph{``the more they bought, the more unique their NFT would be''}. The artwork unfolded through rules of supply and demand, competition, and reward, blurring the line between artistic composition and economic system. Similarly, Pak's earlier collection \emph{X} (2020) consisted of ``infinite editions'' available for only one day, such that scarcity was determined by time and participation rather than a predefined edition size. In all these cases, what constitutes the artwork is not a static file but the entire process governed by protocol: the contracts, algorithms, and user interactions that generate the final outputs. This aligns with Burnham's notion that in system-based art \emph{``all phases of the life cycle of a system are relevant''} and there is \emph{``no end product that is primarily visual''} \cite{burnham1968systems}. Pak's art finds its form in the living system of relationships and behaviors, not in a singular immutable image.

\subsection{Autonomous governance for open-ended control}
Many of Pak's projects foreground autonomous processes and decentralized control, resonating with concepts of generative autonomy in both blockchain and artificial life. On the blockchain, smart contracts act as self-governing \emph{``digital physics''}---immutable rule sets that execute without centralized oversight \cite{ludens2022Autonomous}. This allows artworks to behave like autonomous entities. For example, \emph{Merge} (2021) implemented a self-contained rule whereby all purchased units automatically ``merged'' into singular tokens in each wallet at the close of the sale, without any manual intervention. The final form emerged from the contract's code and collective user actions, illustrating open-ended control dictated by an algorithm. Pak often cedes such control to the system itself or the community, allowing outcomes beyond a single author's full prediction.

This ethos connects to emerging \emph{``autonomous worlds''} \cite{ludens2022Autonomous} in crypto art, and is exemplified by projects like Primavera De Filippi's Plantoid, an art piece that lives on a DAO and \emph{``reproduces itself''} via smart contract governance. In Plantoid, people who hold ``seed'' NFTs fund and vote on the creation of new instances, effectively making the artwork a self-propagating entity. Pak's work incorporates similar principles of shared or automated governance. In \emph{Censored} (2022), the one-of-one \emph{Clock} NFT was auctioned not to an individual but to AssangeDAO, a collective of over 10,000 people pooling funds to bid. A decentralized community thus became the owner, blurring the line between patron and curator through collective governance. Moreover, Pak's \emph{burn.art} platform runs as an ongoing autonomous mechanism: anyone can send an NFT to be burned at any time in exchange for \$ASH, which in turn can be used in future projects. This \emph{``never-ending `Buy--Burn' game''} can continue indefinitely by design, illustrating how the artwork's evolution is handed over to participant actions and blockchain logic. Such autonomous, perpetual systems align with the idea of art as an ``unstoppable'' process---a \emph{``complex living system''} on chain that \emph{``no single entity can halt''} \cite{Hu2024Speculating}. Pak's protocol art thus moves toward open-endedness, where creation and control are distributed across code and community rather than centralized in the artist.

\subsection{Distributed agency and communal authorship}
Pak's protocol-based artworks distribute creative agency among many participants, inviting communal authorship of the art. This follows a broader trend in generative and participatory art where creators \emph{``collaborate with users---either collectors or other artists---to distribute control,''} effectively denying any single central author. In Pak's projects, collectors and players are not passive owners but active co-creators of meaning and content. \emph{Lost Poets} (2021) makes this clear: after minting their AI-generated Poet NFTs, collectors were empowered to name their poets and even invent backstories or verses for them, literally inscribing personal creativity into the work's metadata. As Pak's team noted, the \emph{``discoverers of this civilization will shape it,''} meaning the community of holders defines the cultural narrative of the piece. Indeed, many participants wrote original poems and titles for their Lost Poets tokens, blending their own literary voice with Pak's conceptual framework.

Similarly, \emph{Censored} (2022) turned its audience into co-authors. During the 48-hour open edition, nearly 30,000 messages were tokenized by users and added to the Censored collection. Each participant essentially created a piece of the overall artwork by embedding a personal message on the blockchain. The final collection is thus a mosaic of voices from the community (``Pak \& Assange \& You,'' as the creators billed it), rather than a singular narrative. Even \emph{Merge} can be seen as distributing authorship: almost 30,000 buyers collectively determined the supply and composition of that work's outcome, making the \emph{``most expensive NFT''} \cite{Censored2022} a product of crowd dynamics as much as Pak's initial concept. This radical dispersal of agency recalls Umberto Eco's idea of the ``open work,'' extended into the tokenized realm---the artwork is an evolving network of contributions. By building frameworks that require audience input to complete the work, Pak positions the community as co-creators---enacting what Zeilinger terms \emph{``structures of belonging''} beyond mere ownership \cite{Zeilinger2023Structures}. The artistic authorship becomes plural and emergent from the group, rather than originating solely from the individual artist.

\subsection{Temporal dynamism and lifecycle aesthetics}
Pak's works are not static snapshots; they unfold over time, embracing change, ephemerality, and defined life cycles as core aesthetic elements. In classical systems art, the \emph{``consistency of a system may be altered in time and space''} and all stages of its life cycle become material for the artwork. Pak's projects exemplify this temporal dynamism. \emph{Lost Poets} (2021) was structured as a narrative in multiple acts---from the initial sale of ``Pages,'' to a delayed Reveal where pages transformed into Poet NFTs, through daily drops of special ``Origin'' Poets, and onward towards a planned ``Twist'' and final Epilogue after 365 days. The piece lives and evolves across an entire year, its full meaning only emerging gradually as collectors engage over time. This built-in lifecycle---birth (minting), maturation (naming and interacting), and an end-state after one year (when burning for \$ASH is enabled)---turns time into an artistic medium.

Other Pak works similarly rely on temporal structure. \emph{Censored}'s \emph{Clock} is explicitly a time-based artwork: a dynamic counter tethered to Assange's real-world status---it tallied his days in prison until his release in June 2024, at which point it began counting days of freedom. The open edition portion of Censored was also time-limited to 48 hours, highlighting the urgency and performative moment of the audience's participation. In \emph{X} (2020), each NFT ``moment'' was only available during a one-day window, making \emph{``one single day of infinite existence''} the defining limit of each edition. Scarcity and content in X were thus time-contingent: the fewer people who acted within the day, the fewer copies existed. Even Pak's market dynamics often have temporal phases (for instance, The Fungible's escalating price tiers each day). By designing pieces with evolving states and deadlines, Pak \emph{``recreates the cycle of life''} in digital form, acknowledging that in a systemic artwork, impermanence and transformation are not byproducts but the very substance of the art.

\subsection{Economy-driven engagement}
Pak's projects deliberately intertwine economic mechanisms with artistic interaction, turning market dynamics into part of the artistic experience. In many cases, the engagement of the audience is driven by financial gameplay---bidding wars, token burns, price competition---such that market participation becomes a form of performance. The Sotheby's sale \emph{The Fungible} is a paradigmatic example: it deployed a complex auction structure with open editions, surprise price hikes, and contests for unique rewards, effectively a ``gamified'' economic performance. This not only generated approximately \$16.8 million in sales, but also created a frenzy in which \emph{``more than 3,000 individuals decided to join,''} drawn in by the competitive game itself. The mechanics of scarcity and reward (e.g., top buyers receiving special NFTs) were designed to incentivize maximal participation. Here the market behavior was the art: once \emph{``the game and competition incentives disappear,''} what remains are tokens whose value was defined by that performative context. Pak thus highlights, even cynically, the extent to which economic structures confer meaning in crypto art \cite{Poposki2025Critical}.

In Pak's oeuvre, the act of buying or transacting becomes a critical ritual and medium. \emph{The Title} (2021) consisted of identical images sold at different prices, making price itself the distinguishing feature of the artwork. One edition titled ``Unsold'' was listed for \$1 million and intentionally left unsold---a conceptual gesture underlining how value in art can be a product of hype and perception. This recalls Yves Klein's 1957 exhibition of identical blue canvases sold at varying prices; as Klein observed, \emph{``the price…legitimately changes the experience of the work''} despite no physical difference. Similarly, Pak has argued that an NFT is like a currency---judging it by the image alone is as foolish as picking a banknote. By merging art and economics, Pak invites collectors to reflect on their own motivations---are they driven by aesthetic appreciation or speculative gain? The economy around the artwork becomes integral to its narrative, effectively turning collectors into players in a financial drama. In Pak's protocol art, market engagement is not a byproduct but a core feature that actively shapes the work itself.

\subsection{Poetic message embedding in interaction rituals}
Beyond the technical and economic layers, Pak's works often carry metaphorical or poetic messages that are realized through the audience's interactive rituals. In other words, the process a participant must follow is itself laden with symbolic meaning. Critic A.V. Marraccini notes that Pak treats the NFT medium as \emph{``a way to engage audiences in participatory conceptual art''}---the concepts emerge through what the viewers or users are made to do. For instance, burn.art turns the destructive act of burning NFTs into a creative ritual. The platform's mantra \emph{``Burn art to get ashes to get art to burn art''} is a circular poem in itself, invoking themes of death and rebirth. Participants enact a cycle of sacrifice and renewal: by destroying one token, they generate another (the \$ASH token), which can then be spent on new art or even burned again. This cyclical interaction is a performative allegory about value and transformation---literally illustrating creation-through-destruction in a way that words alone could not.

Pak's collaborations also embed messages in their structure. \emph{Censored} (2022), created with Julian Assange, is explicitly about free expression and censorship. The project had two parts---one a dynamic single edition (Clock) counting Assange's days behind bars, and the other an open invitation for people to speak. By telling users that \emph{Censored} is ``about you'' and allowing them to tokenize any message they wish, Pak made the audience's personal expression the core of the piece. The very name \emph{Censored} reminds participants that their recorded messages might be obscured or blacked out, forcing reflection on the power of speech under observation. In \emph{Lost Poets}, Pak evokes Jorge Luis Borges's Library of Babel: an infinite library of all possible texts. The project \emph{``enlists participants to imagine''} the lost contents of that library by naming their Poet NFTs and eventually inscribing words onto them. The ritual of naming and writing thus becomes a meditation on how meaning is generated---each user's creative act is a fragment of a larger, hidden poem. Pak uses these game-like interactions not just as gimmicks but to convey ideas: the audience realizes the ``poetic'' concept by performing it. This approach recalls aspects of Fluxus or conceptual art, where simple actions (publishing a message, paying an exorbitant sum) become symbolic acts. The blockchain in Pak's work provides a global, participatory stage for such rituals, ensuring that each interaction---each burn, each tokenized word---is both an artwork and a narrative gesture in Pak's conceptual universe.

\subsection{Interoperability enabling composability for emergence}
Finally, Pak's protocol-based approach leverages the interoperability of blockchain tokens to compose complex emergent ecosystems that transcend any single artwork. In blockchain terms, assets and smart contracts are composable: they can interact, integrate, or build upon each other permissionlessly. Pak exploits this composability to create an interlinked network of projects. The clearest example is the \$ASH token ecosystem. After \emph{The Fungible}, Pak introduced \$ASH as a cryptocurrency earned by burning NFTs on burn.art. This token was not confined to one piece, but became a connective thread between works---a meta-artwork in its own right. Owners of \$ASH were rewarded in subsequent projects: before launching \emph{Lost Poets}, Pak took a snapshot and airdropped 7,586 Pages to collectors holding at least 25 \$ASH, thus bridging the burn.art economy with the new literary NFT game. Conversely, once Lost Poets concluded, those Poet NFTs could be burned for \$ASH, feeding value back into the burn.art system. In this way, Pak created a feedback loop: one artwork's output becomes another's input. The emergent whole is an ecosystem of NFTs and tokens co-evolving across Pak's oeuvre.

This interoperability enables creative ``combinatorics'' that yield unforeseen outcomes. Smart contract platforms allow \emph{``multiple `worlds' to intertwine,''} as researchers note, forming a \emph{``persistent, yet sufficiently complex environment''} where artworks can coexist. Pak's art inhabits this multiverse. A token like \$ASH grants access to exclusive drops and collaborations (a composable utility), while the act of burning or merging tokens can spawn new forms and communities. By giving participants portable assets that carry value and function across different contexts, Pak encourages emergent behavior. Collectors speculate, strategize, and invent new use-cases (for example, pooling \$ASH or devising burn strategies), effectively participating in the evolution of the creative system. As one commentator observed, Pak's project proved that on blockchain \emph{``individuals can exercise personal control over value, create niche ecosystems…without the sluggish constraints of fiat,''} demonstrating the \emph{``power of creative tokenomics''} in art.

\section{Discussion}

\subsection{Characteristics of Pak's Protocol Artworks}

This study's exploration of Pak's protocol artworks reveals seven interlinked characteristics that distinguish protocol art as a practice. \textbf{(C1)} These works are fundamentally system-centric rather than object-centric. Their artistic meaning emerges from dynamic interconnectedness via protocols instead of static artifacts. \textbf{(C2)} As Galloway claimed \textit{``protocol is how control exists after decentralization''} \cite{galloway2001protocol}, protocol arts employ autonomous governance for open-ended control, often via self-executing code or decentralized mechanisms, so that the artwork's evolution is not predetermined by a single author but can continue to unfold indefinitely. \textbf{(C3)} This leads to distributed agency and communal authorship, as creative influence is spread across many participants and technical actants rather than residing in one creator, aligning with the idea that agency (and thus authorship) is shared among human and non-human actors in a network. \textbf{(C4)} Moreover, these artworks exhibit temporal dynamism and lifecycle aesthetics: they develop through time, with phases of growth or change designed into the experience, so that process and feedback become central to their aesthetic. \textbf{(C5)} Engagement with such works is often economic-driven, integrating incentive structures (for example, token rewards or market dynamics) that actively shape how audiences interact---a phenomenon noted by art scholars observing the immediate economic incentives at play in crypto art. \textbf{(C6)} At the same time, there is a deliberate embedding of poetic messages in interaction rituals: the very actions participants perform (trading, voting, combining elements, etc.) are imbued with symbolic meaning, turning user interactions into ritualized performances that carry the artwork's conceptual message. \textbf{(C7)} Finally, protocol artworks leverage interoperability enabling composability for emergence, meaning they are designed to plug into larger ecosystems of code and community; like modular components, they can be reconfigured or linked with other protocols, allowing unexpected behaviors and creative outcomes to emerge from these combinations. Collectively, these seven characteristics illustrate how protocol art shifts the locus of art-making from singular objects to evolving systems of interaction, value and meaning.

By situating these findings in broader intellectual contexts, we can appreciate protocol art as part of an interdisciplinary conceptual lineage. The system-centric focus strongly resonates with complex systems theory and the ``systems aesthetics'' tradition in art: as Jack Burnham observed, modern art increasingly \textit{``does not reside in material entities, but in relations between people and between people and the components of their environment''}. In protocol art, the artwork is essentially a complex adaptive system \cite{buckley2017society}---with inputs, outputs, and feedback loops---which accords with theoretical models where adaptive behavior and emergent order arise from networked interactions over time. Its open-ended governance and temporal evolution connect to the field of artificial life, which seeks systems that continually produce novelty rather than reaching a fixed end state. This open-endedness \cite{Soros2017Openendednessb}, coupled with deliberate feedback mechanisms, echoes the temporal dynamics of complex adaptive systems studied in complexity science, reinforcing how unpredictable, ongoing change becomes an aesthetic virtue. Meanwhile, the ethos of distributed agency and communal authorship in these works finds precedent in media theory and anthropology: for instance, actor-network theory contends that creative agency is distributed across a web of human and non-human actants, not concentrated in an autonomous individual. Similarly, in interactive and participatory art literatures, authorship is often viewed as an emergent, collective process---the community of participants effectively co-creates and continually reshapes the piece \cite{bishop2023artificial}. The inclusion of economic-driven engagement situates protocol art in dialogue with social computing and crypto-economic systems: just as Web3 platforms rely on token-based incentives to drive user participation and loyalty, protocol artworks harness economic game dynamics as artistic material, blurring the line between aesthetic experience and market behavior. This integration of economic and social incentives also invokes evolutionary game theory within complex systems, where value-based choices influence the system's trajectory. Furthermore, the poetic messaging through interaction rituals can be interpreted via performance studies and ritual theory in media. Even routine or rule-bound interactions carry expressive and symbolic weight---indeed, ritualized acts are \textit{``anything but purposeless''} and can \textit{``constitute social reality''} through shared symbolic action. In protocol art, the choreography of user interactions (such as collective decision-making or repetitive transactions) functions as a form of narrative or commentary, akin to a ceremonial performance that conveys meaning beyond its practical function. Finally, the principle of interoperability and composability aligns with discussions in software and platform studies about modular design and emergent behavior. In blockchain-based art, for example, smart contracts are conceived as interoperable building blocks, combinable like Lego pieces to create novel structures \cite{sun2024smart}. This composability not only reflects a technical design philosophy but also fosters creative emergence: new artistic forms and communities can arise by linking protocols together, much as complex behaviors emerge when simple units interact in networked ecologies. Linking these characteristics to established frameworks thus shows that protocol art is not an isolated novelty but converges with long-running threads in complexity science, new media art, and socio-technical theory, reframing them in a unique artistic context.

\subsection{Distinction from Adjacent Movements}
Pak's oeuvre inherits strategies from Conceptual, Generative, Participatory, Interactive, and Performance art, yet the blockchain protocol gives those strategies a qualitatively different reach. What follows parses each genealogy and shows where Pak's ``protocol art'' outruns its nearest relatives.

\subsubsection{Beyond Conceptual Art}

\emph{``The idea becomes the machine that makes the art,''} wrote Sol LeWitt in 1967 \cite{lewitt1967paragraphs}---a formulation that conceptual art could only fulfill metaphorically, since LeWitt's wall drawings still required human execution and institutional presentation.
Pak's protocol art literalizes this aspiration: the smart contract \emph{is} the machine, and the idea---encoded once as immutable bytecode---executes itself without further authorial intervention.
Joseph Kosuth's \textit{One and Three Chairs} (1965) established that an idea can eclipse the physical object when reframed through language and documentation.
Pak embraces the primacy of ideas but transports them into a rule-executing substrate.
\emph{The Title} (2021), whose only variables are price and on-chain title, echoes Kosuth's linguistic gambit, yet its value is continuously arbitrated by decentralized markets rather than by academic discourse or institutional framing.
Because the smart contract is immutable, its ``statement'' cannot be re-contextualized by curators; it is enacted in perpetuity by the network.
Thus Pak's work is not merely an illustration of a concept but a permanently operational logic that strangers must literally transact with.
The artwork lives as executable code---a self-verifying proposition---rather than a documented proposition that awaits institutional re-presentation.

\subsubsection{Beyond Generative Art}

Early computer artists such as Vera Molnár treated algorithms as mechanical aides that output static images; even later real-time generative pieces, like Casey Reas's \textit{Process} series, run autonomously but remain visually contained.
Pak's generativity is socio-economic: the code sets only the initial fitness landscape, while thousands of human agents supply the variation and selection pressures that drive the system forward.
In \emph{Merge} (2021), the merging rule is trivial---add two integers---yet the competitive hunt for ``Alpha Mass'' steers the macro-form in ways no solo algorithm could predetermine.
Where classic generative art is a dialogue between artist and machine, Pak's field is a triadic ecology of artist, code, and market actors; the emergent aesthetic includes price curves, social memes, and token topologies---materials that lie outside the scope of conventional algorithmic art.

\subsubsection{Beyond Participatory / Crowdsourcing Art}

Relational-aesthetic projects of the 1990s (Bourriaud) invited viewers to co-produce meaning through convivial encounters, but those encounters were typically ephemeral and locally bounded.
Pak's participation is \emph{ledger-bound}.
In \emph{Lost Poets} (2021--22), each name or verse a collector inscribes becomes an indelible block in a public archive---a form of distributed authorship impossible in earlier participatory works whose traces remained in wall labels or catalogs.
Crucially, participation is also economically weighted: burning a Page to earn words entails an explicit cost, so creative decisions carry financial risk.
This introduces game-theoretic depth absent from classic crowdsourcing art (e.g., MTurk drawings or open Wikis), aligning Pak's practice with emerging literatures on cryptoeconomic design rather than with social-practice documentation alone.
Moreover, \emph{Censored} (2022) demonstrates that protocol-based participation can transcend the art world entirely. AssangeDAO---a collective of over 10,000 members---pooled 16,593 ETH (approximately \$52.8 million) to acquire the \emph{Clock} NFT, with proceeds directed to Julian Assange's legal defense via the Wau Holland Foundation. When Assange was released in June 2024, the participation had demonstrably contributed to a real-world political outcome---a reach no gallery-bound participatory artwork has achieved.

\subsubsection{Beyond Interactive Art}

Interactive art from Myron Krueger's \textit{Videoplace} to Rafael Lozano-Hemmer's public-space works invites real-time feedback loops, but those loops terminate once the exhibition ends.
Protocol art designs interaction into on-chain fact.
When a collector merges two \emph{Merge} tokens or burns an NFT for \$ASH, the blockchain records a state change that affects every future viewer and holder.
Interaction ceases to be a momentary spectacle and becomes infrastructure---a permanent mutation in the artwork's ontology. Each transaction---a merge, a burn, a tokenized message in \emph{Censored}---is an irreversible on-chain fact, not a fleeting gesture that vanishes when the gallery closes.
Moreover, protocol interaction is asynchronous and scalable: millions of transactions are handled by the Ethereum network itself, so participation is bounded only by the scale of the chain rather than by gallery capacity or curatorial gatekeeping.

\subsubsection{Beyond Performance Art}

Performance art often foregrounds the performative encounter between artist and participant---think of Yoko Ono's \textit{Cut Piece} (1964) \cite{Ono1964Cut}, in which audience members were invited to cut away the artist's clothing, making collective action the medium.
Protocol art extends this logic beyond the gallery and across the entire internet: anyone, anywhere, can permissionlessly participate in the work by signing a transaction, burning a token, or inscribing a message---no invitation, no curator, no physical co-presence required.
Where Ono's performance was bounded by a single room and a single sitting, \emph{Censored} (2022) enrolled nearly 30,000 participants across two days and multiple continents, each contributing a locked message that became part of an enduring on-chain chorus.
The protocol replaces the artist's body as the performing agent: it executes its rules indefinitely, delegating durational endurance to automated smart-contract logic and DAO stewardship rather than to flesh and presence.
Thus the locus of performance shifts from a singular body in time to a permissionless, planetary-scale system in perpetuity.

\subsection{Limitations and Future Work}

Our intensive case study of Pak's nine landmark protocol artworks demonstrates how smart-contract logic can serve as an artistic medium while providing foundational vocabulary for the emerging field. By focusing on the paradigmatic creator of protocol art, we have articulated seven characteristics that can now serve as analytical lenses for future comparative research.

This study does not claim that all protocol art exhibits the seven characteristics we identify; such a claim would require cross-artist sampling inappropriate to our exploratory aims. Instead, we offer a \emph{generative} contribution: a vocabulary and analytical framework derived from close engagement with protocol art's most influential practitioner. These characteristics function as what Blumer termed \emph{sensitizing concepts}---interpretive devices that orient attention without predetermining findings---that future researchers can apply, adapt, or reject when analyzing other artists \cite{Blumer1954}. The value of our contribution lies not in statistical generalization but in conceptual articulation: naming and describing phenomena that lacked prior theorization. Whether terra0, Sarah Friend, or Rhea Myers exhibit similar characteristics is an empirical question our framework now makes askable. Skeptics have questioned whether NFTs substantially modify art-world dynamics at all \cite{Radermecker2023Questioning}, making such comparative work all the more pressing. We anticipate that some characteristics (e.g., system-centric composition, autonomous governance) will prove broadly applicable to protocol art as a genre, while others (e.g., economy-driven engagement at Pak's scale) may be specific to commercially-oriented practice. Comparative research is needed to distinguish genre-defining features from artist-specific idiosyncrasies.

At the same time, this focus exposes the field's present fragmentation: there is still no shared vocabulary, taxonomy, or methodological toolkit robust enough to account for the full diversity of protocol-driven practice. We therefore call for a concerted, interdisciplinary effort to systematize knowledge around protocol art. Concretely, future work should
\begin{itemize}
\item map a wider corpus of artists and projects to test and refine the analytical lenses introduced here;
\item establish common descriptors for protocol mechanics (e.g., governance rules, token dynamics, temporal constraints) that are comparable across cases;
\item develop metrics---both qualitative and on-chain quantitative---for assessing participation, emergent behavior, and sociocultural impact;
\item integrate perspectives from art history, HCI, STS, economics, ethics \cite{Calvo2024Cryptoart}, and legal studies to ground a theory of ``protocolism'' that distinguishes it from adjacent genres such as generative art and instruction-based conceptual art.
\end{itemize}

\section{Conclusion}
Our case study of Pak's nine landmark protocol artworks reveals seven distinct characteristics of protocol art, helping advance both its definition and our understanding of this emerging genre. Protocol art remains an emerging creative domain that lacks a comprehensive theoretical framework. While ``protocolism'' conceptualizes protocols as an artistic medium, the field needs more rigorous systematization of its concepts and aesthetics. Current discourse around blockchain-based art has emphasized market dynamics and hype over conceptual understanding \cite{Poposki2024Corpus}. This rapid evolution has created a significant gap between practice and theory---one that requires careful critical reflection and theoretical development to properly ground the field.

\balance
\bibliographystyle{ACM-Reference-Format}
\bibliography{reference}

\clearpage 

\appendix

\section{Pak's Protocol Artworks}

This appendix contains collected information analyzing Pak's protocol art through eight analytical lenses summarized and derived from ChatGPT-o3 Deep Research: (1) Concept, (2) System-mechanism dynamics, (3) Participatory interaction, (4) Artist's control, (5) Collective emergent behavior, (6) Poetic meaning-making, (7) Key statistics, and (8) Annotation.

\subsection{X (2020)}

\begin{figure}[ht]
    \centering
    \includegraphics[width=0.9\linewidth]{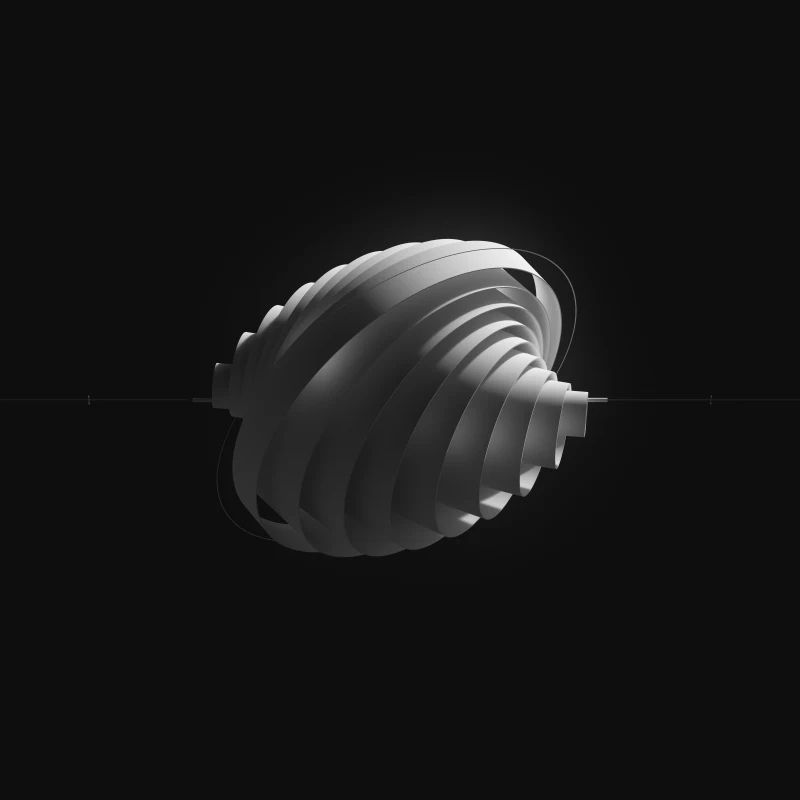}
    \Description{A 3D-rendered sculptural form composed of layered horizontal slices arranged into an ellipsoidal shell, rendered in grayscale with soft lighting against a black background, bisected by a thin horizontal line.}
    \caption{\textit{The Wave}, edition of 3, from Pak's \textit{X} collection (2020). Thirteen of the 15 pieces in \textit{X} shared a time-based open-edition format where scarcity was determined by a 24-hour minting window rather than a fixed supply; the remaining two were one-of-one auctions.}
    \label{fig:x}
\end{figure}

\paragraph{Concept} \emph{X} was Pak's pioneering experiment in defining digital scarcity through time rather than supply. Launched in August 2020 as an open-edition NFT sale on Nifty Gateway, it replaced fixed edition limits with a 24-hour minting window. This inversion of the usual rarity model made the concept of \emph{X} about temporal urgency---the art's value derived from the moment of its creation rather than a predetermined quantity. In doing so, Pak introduced a novel protocol-native idea: that the conditions of a sale (time constraints and participation) can themselves be the conceptual core of the artwork, shifting focus from image scarcity to the shared time-bound experience.

\paragraph{System Mechanism} The \textit{X} project consisted of 15 NFT pieces (13 open edition works and 2 one-of-one auctions) released simultaneously. Each open edition piece had no cap on mint count; instead, a 24-hour window was the sole limit. This scarcity mechanism \emph{``based on time rather than volume''} was unprecedented at launch. Once the window closed, each piece's final edition size was fixed forever by the number of mints in that period. Technically, Pak's smart contract minted new tokens on demand during the sale, then refused further minting after the deadline, enforcing time-limited creation. The dynamics encouraged rapid uptake: as time dwindled, indecisive collectors had to commit or miss out. In parallel, two unique NFTs were auctioned, adding a traditional scarcity element. The interplay of these formats (open vs. fixed quantity) within \textit{X} highlighted the NFT's programmability---scarcity became a flexible parameter of the artwork. Post-drop, the open edition pieces entered the secondary market with widely varying supply counts, testing how market dynamics would value an NFT explicitly designed without a fixed edition size.

\paragraph{Participatory Interaction} Collectors engaged with \emph{X} by acquiring the NFTs during a strictly limited timeframe, effectively collaborating in a one-day event rather than competing for scarce editions. Anyone could mint as many as desired within the 24-hour window, turning the drop into a collective performance synchronized in time. This format, then-unusual, invited a broad pool of participants to be part of the work's creation---an approach that soon became a common practice in crypto art. The emphasis was on open participation: the art market's typical frenzy for limited pieces was replaced by an inclusive but fleeting opportunity, underscoring the role of community presence and timing in the piece.

\paragraph{Artist's Control} After \emph{X} was launched, Pak's direct control was minimal---the rules were preset in the smart contract, and the outcome (how many tokens minted) was determined by the community within the allowed time. Pak's role was chiefly in conceptualizing and coding the parameters; once the clock started, the process was autonomous and irreversible, with Pak acting more as instigator than controller. This ceding of control to the protocol is central to the work's philosophy: the artist designed the playground but did not interfere with how many editions were ultimately created or who obtained them. The result is that \emph{X} lives on as a completed event etched on-chain, with Pak's influence embedded in the mechanism but not exerted in real-time during or after the sale.

\paragraph{Collective Emergent Behavior} \emph{X} attracted a broad swath of NFT collectors due to its open nature, with every buyer during that day becoming part of the work's narrative. The final mint counts effectively recorded the number of participants---a transparent metric of engagement---and in total, thousands of NFTs were minted, involving a large community. Rather than fostering a competitive race for a single token, \emph{X} created a shared experience: collectors knew the availability was equal for all but also ticking away, which galvanized social media buzz and a sense of camaraderie among those ``in the moment.'' This collective temporal convergence---everyone minting in the same 24-hour span---became a form of emergent community behavior, reinforcing Pak's notion that the artwork was as much about the crowd's response as the digital objects themselves.

\paragraph{Poetic Meaning-Making} \emph{X} transforms the notion of an artwork into a time-bound performance, offering a poetic commentary on how value and meaning in digital art are shaped by temporal and social parameters. The piece demonstrates that an NFT's significance can lie not just in its image but in the circumstances of its creation---here, a collective moment that cannot be replicated. By foregrounding the protocol (the 24-hour clock) over the visual content, Pak underscored themes of ephemerality and community: the fact that \emph{X} could only come into being through synchronized action imbues it with a narrative of shared presence. Ultimately, \emph{X} invites reflection on an alternate mode of art valuation: one where scarcity is a function of time and collective presence rather than material limitation, thereby poetically aligning the artwork with the ethos of a decentralized, time-sensitive digital culture.
\paragraph{Key Statistics} Released 28 August 2020 on Nifty Gateway, the \textit{X} collection ultimately comprised 15 artworks with a total of 63 editions minted (across 13 open works and 2 one-of-ones). Over 24 hours, collectors minted as few as 3 and at most 8 copies per open-edition piece, totaling only 63 pieces---far fewer than the theoretical infinity, underscoring the experiment's success. The highest known sale from \textit{X} was the unique \textit{The Void}, which was auctioned to a single collector (price undisclosed publicly), while aggregate primary sales for \textit{X} exceeded six figures in USD terms (each open edition priced in the low hundreds of dollars). This series marked Pak's first major NFT drop, establishing the artist's reputation for novel mechanics in crypto art.

\paragraph{Annotation} \emph{X} is the foundational case for \textbf{C1} (system-centric composition) and \textbf{C4} (temporal dynamism): the artwork has no fixed edition size---its supply is a function of collective behavior within a 24-hour window, making the protocol's time constraint, rather than any visual property, the locus of artistic meaning. What makes this particularly significant for digital art is the inversion of the scarcity paradigm: while artists can trivially copy digital files, market demand cannot be copied, and Pak exploits this asymmetry by substituting a temporal mechanism for the conventional predetermination of edition count. Scarcity in \emph{X} is not declared by the artist but \emph{discovered} through market-driven behavior---a limited window generates a limited edition without any authorial fiat over supply. This absence of predetermination is conceptually radical: it demonstrates that the scarcity of digital artwork need not be imposed from above but can emerge from the structure of participation itself. \emph{X} thus initiates \textbf{C5} (economy-driven engagement) by turning the act of purchasing into the generative act itself, carrying a poetic meaning of market dynamism in which economically driven engagement literally constitutes the work. Notably, \emph{X} shows minimal \textbf{C2} (autonomous governance) and no \textbf{C7} (composability): it is a self-contained event with no token mechanics that bridge to later works. As Pak's earliest NFT project, \emph{X} establishes the core intuition---scarcity as a protocol parameter rather than an artist's decree---that all subsequent works elaborate.

\paragraph{Information}

\begin{itemize}
\item 
Launch Website: \url{https://niftygateway.com/collections/pak}
\item 
Release Date: Aug 28, 2020
\item 
OpenSea: \url{https://opensea.io/collection/x-by-pak}
\item 
Contract Address: \\0x99b546a19cc1ec8ec9a6ce781a237ddb642dda77 (ERC721)
\end{itemize}

\subsection{The Title (2021)}
\begin{figure}[ht]
    \centering
        \includegraphics[width=0.9\linewidth]{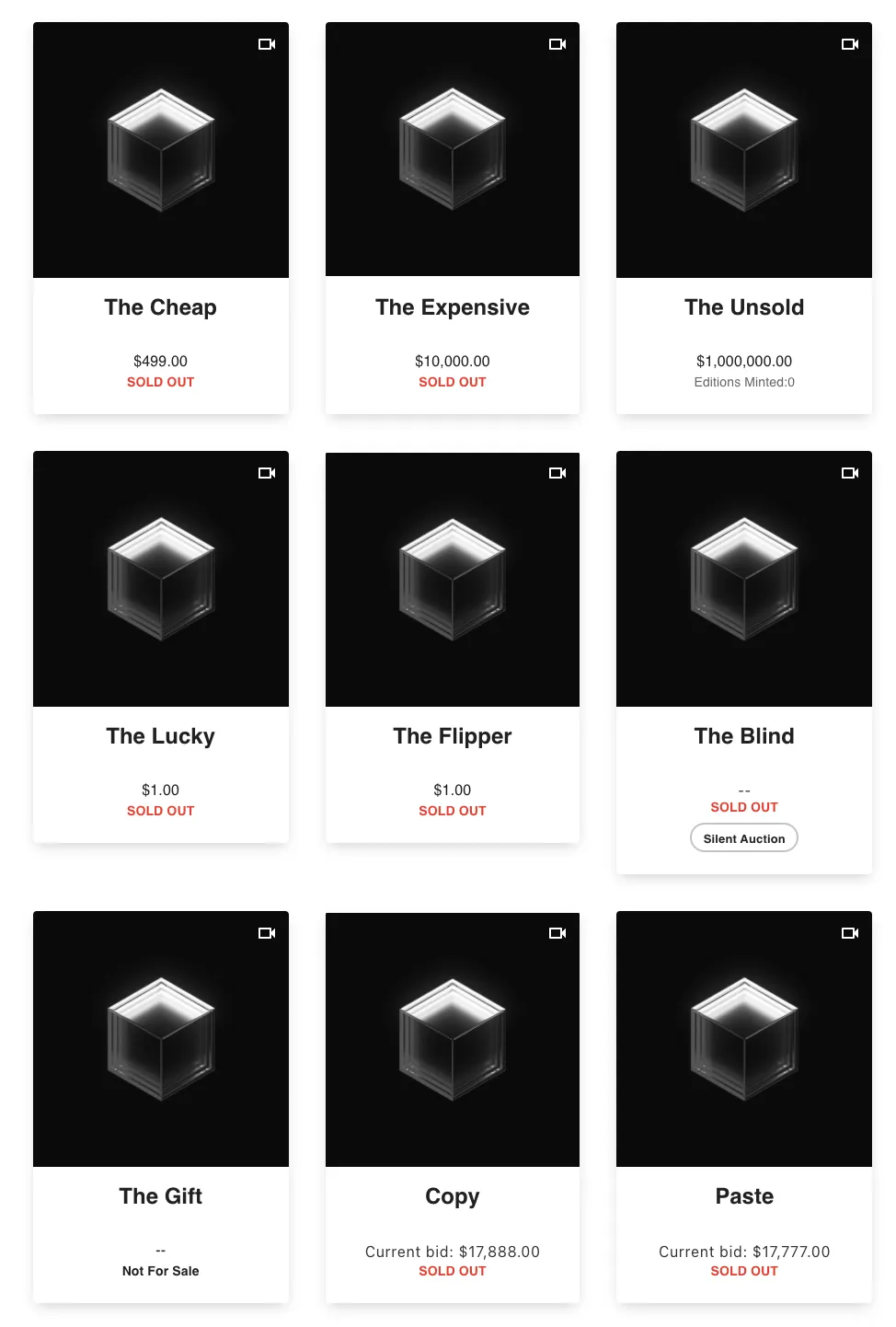}
    \Description{A screenshot of the Nifty Gateway marketplace showing a three-by-three grid of nine NFT listings, each displaying the same rotating translucent 3D cube but with different titles and prices: The Cheap (499 dollars), The Expensive (10,000 dollars), The Unsold (1,000,000 dollars, zero editions minted), The Lucky (1 dollar), The Flipper (1 dollar), The Blind (silent auction), The Gift (not for sale), Copy (17,888 dollars, sold out), and Paste (17,777 dollars, sold out).}
    \caption{Marketplace listing for \textit{The Title} by Pak (2021). All nine NFTs share an identical rotating cube visual; only the on-chain title, price, and distribution method distinguish each piece---foregrounding naming and context over visual content as determinants of value.}
    \label{fig:thetitle}
\end{figure}

\paragraph{Concept} \emph{The Title} (January 2021) is a conceptual NFT series that interrogates the nature of value and ownership in crypto art. Pak created nine virtually identical digital objects---all a rotating translucent cube---but each NFT was given a different title and method of distribution \cite{e2021MuratPaks}. By selling the same image at different prices and under labels like \emph{The Cheap}, \emph{The Expensive}, \emph{The Unsold}, etc., Pak demonstrated that the perceived value of an NFT lies not in the visual content but in contextual factors such as scarcity, nomenclature, and narrative. The artwork's core concept is essentially a self-referential critique of the NFT medium: \emph{The Title} asks what it is we truly buy when we purchase an NFT, positing that the token's title and associated story can override the image in defining its essence.

\paragraph{System Mechanism} Under the hood, \emph{The Title} utilized a clever technical setup: all nine NFTs pointed to the exact same IPFS-hosted media file of the spinning cube. Despite this shared imagery, the smart contract distinguished each token by metadata---notably the title and edition info---which conferred each piece's unique identity and implied value. This meant that ownership of any one token was effectively ownership of the same underlying image, forcing a paradox: collectors held different tokens but ``shared'' the art content. The dynamics of this system laid bare the role of metadata and token provenance in NFT valuation; questions arose such as whether the holder of ``The Expensive'' had more claim to the art than the holder of ``\emph{The Cheap},'' even though the visual asset was identical. By structuring the NFTs in this way, Pak's smart contract became a philosophical device, highlighting how blockchain metadata and distribution rules can generate distinct meanings and hierarchies around an otherwise fungible digital image.

\paragraph{Participatory Interaction} Collectors engaged with \emph{The Title} through a gamified multi-format sale. Each of the nine NFTs was obtained via a different mechanism---some were open editions (available to anyone for a limited time), others were limited editions at set prices, one was sold via blind auction, one via a standard auction, and one (``\emph{The Gift}'') was not sold at all but given away \cite{e2021MuratPaks}. This design meant that the audience had to navigate various modes of acquisition, effectively turning the act of collecting into part of the artwork's performative narrative. By strategically involving buyers in auctions, giveaways, and timed drops, Pak cultivated a community-wide discourse and excitement around the project. The process of obtaining the pieces became a participatory spectacle of its own, reinforcing the idea that the artwork extends beyond the image to include the protocols and social interactions of its distribution.

\paragraph{Artist's Control} Pak's control over \emph{The Title} after release was primarily conceptual and pre-programmed. Once the pieces were distributed through their various channels, Pak did not intervene in the tokens' existence or content---in fact, by anchoring all tokens to a single immutable IPFS image, Pak relinquished any ability to differentiate or alter the visual component for one token without affecting them all. The only lever of control exerted was at inception: defining the titles and scarcity of each edition, thereby scripting the value narrative in advance. After launch, the market of collectors took over; the artist's role shifted to observer of the unfolding debate and trade. This limited post-launch control aligns with the work's intention: it left collectors to ascribe meaning and value among themselves, illustrating Pak's point that the community's perceptions and the immutable smart contract rules (not the artist's hand) ultimately govern an NFT's fate.

\paragraph{Collective Emergent Behavior} \emph{The Title} engaged a diverse group of participants, from high-end collectors willing to pay a premium in auctions to newcomers minting the affordable open edition. The total collector base spanned those nine sub-communities, all linked by the intrigue of the experiment. This project fostered a strong sense of community discussion; buyers compared their acquisitions (proudly identifying as the one who got ``The Expensive'' or the lucky recipient of ``\emph{The Gift}''), and observers debated the merits of each token on social media. Because each token's status (cheap vs. expensive, sold vs. unsold) was transparent on the blockchain, a collective narrative emerged in which the community itself assigned cultural significance to each piece beyond Pak's initial labels. In essence, the collectors became unwitting collaborators in Pak's social experiment, their behaviors---be it speculative flipping of ``\emph{The Cheap}'' or holding ``\emph{The Unsold}'' as a trophy---illustrating how a network of participants can create a rich tapestry of meaning around otherwise identical digital artifacts.

\paragraph{Poetic Meaning-Making} The poetic impact of \emph{The Title} lies in its elegant exposure of art-world conventions transposed to the blockchain. By echoing Yves Klein's 1957 exhibition of identical blue paintings sold at different prices \cite{Blue2020}, Pak placed NFT culture in dialogue with historical avant-garde inquiries into what art really is. \emph{The Title} makes viewers ponder why one iteration of a digital cube should carry more prestige or value than another. It reveals that the poetry of the piece comes from the questions it raises: What is ownership? What do we value in art---the image, the story, or the token that confers bragging rights? The work's broader significance is in demonstrating that blockchain art can critique itself; it's simultaneously a satire and a celebration of the cryptographic medium, showing how scarcity, community perception, and narrative alchemy collectively create meaning in digital art. In sum, \emph{The Title} stands as a meta-artwork that is reflexive about its own value structure, engaging the art discourse on authenticity, authorship, and the economics of the intangible.
\paragraph{Key Statistics} Launched 6 January 2021 on Nifty Gateway, \textit{The Title} comprised 9 works: \textit{The Cheap} (192 editions at \$499 each), \textit{The Expensive} (8 editions at \$10,000 each), \textit{The Unsold} (1 edition, priced so high it remained unsold), \textit{The Blind} (3 editions via silent auction), \textit{The Flipper} (99 editions first-come at a low price), \textit{The Lucky} (3 editions first-come, price \$1 each---effectively a lottery of speed), \textit{Copy} and \textit{Paste} (each 1/1 auctions, paired conceptually), and \textit{The Gift} (3 editions given free to early collectors). Total sales exceeded \$200,000 across the series. Critically, all nine tokens reference one image file, illustrating Pak's point. Sotheby's later noted this collection as a breakthrough in conceptual NFT art, aligning Pak with avant-garde strategies of value interrogation.

\paragraph{Annotation} \emph{The Title} is the purest expression of \textbf{C5} (economy-driven engagement) and \textbf{C6} (poetic message embedding) in the portfolio: by selling identical images at different prices under different names, Pak isolates price and metadata as the sole carriers of meaning, forcing participants to confront the question of whether value resides in the visual object or in the narrative apparatus that frames it. The satirical force of the piece lies in its provocation: since a smart contract requires a unique identifier for each entry, the work asks whether it is the \emph{name} itself that holds primary value---whether the title, not the content, is what collectors truly acquire. This interrogation strikes at the heart of non-fungibility: even identical images, once assigned different names within the contract, become formally distinct, non-fungible things, revealing that the ontological boundary between fungible and non-fungible may hinge on something as thin as a string of metadata. The work thus exemplifies \textbf{C1} (system-centric composition) at its most distilled, since the ``artwork'' is the distribution structure and naming convention itself rather than the rotating cube that all nine tokens share. \emph{The Title} is, in this sense, a work of ``protocol art'' in the strictest definition: the contract \emph{is} the piece, and its poetic message emerges from the nature of the contract's requirements. However, \emph{The Title} is an outlier for \textbf{C3} (distributed agency): collectors cannot alter the work or contribute content---they can only choose which token to acquire. This constraint is itself instructive: it reveals that communal authorship is not essential for protocol art to function as critique, but its absence limits the emergent social dynamics observed in later projects.

\paragraph{Information}

\begin{itemize}
\item 
Launch Website: \url{https://niftygateway.com/collections/thetitle}
\item 
Release Date: Jan 6, 2021
\item 
OpenSea: \url{https://opensea.io/collection/the-title-by-pak}
\item 
Contract Address: \\0x090c53bac270768759c8f4c93151bd1a808a280e (ERC721)
\end{itemize}

\subsection{The Fungible}

\begin{figure}[ht]
    \centering
    \includegraphics[width=0.9\linewidth]{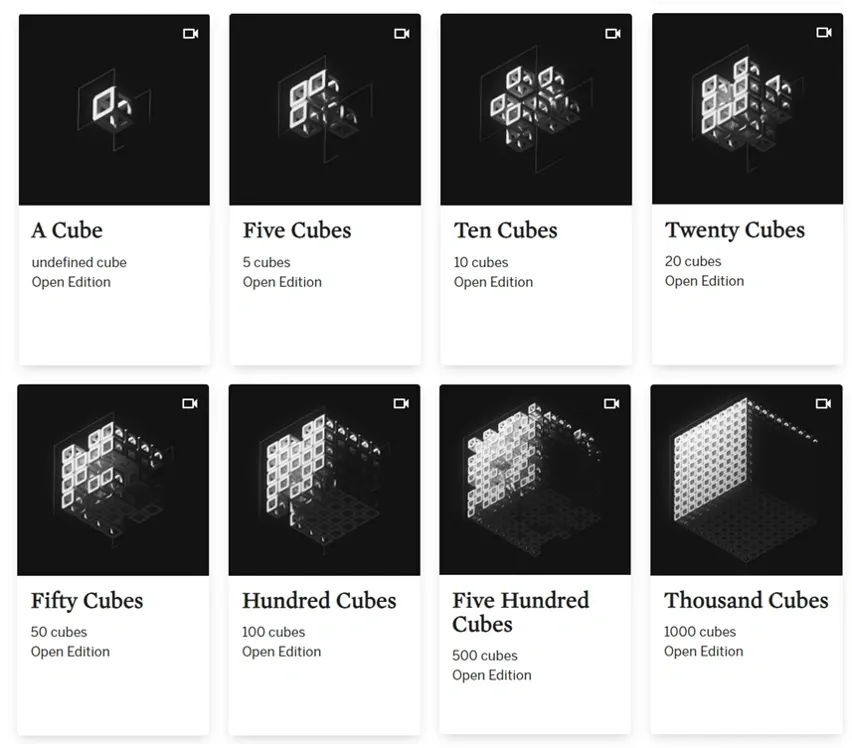}
    \Description{A screenshot of the Nifty Gateway marketplace showing a two-by-four grid of open-edition NFT listings. Each listing displays a different quantity of translucent 3D cubes arranged in grid formations against black backgrounds: A Cube (1 cube), Five Cubes, Ten Cubes, Twenty Cubes, Fifty Cubes, Hundred Cubes, Five Hundred Cubes, and Thousand Cubes. All are labeled Open Edition.}
    \caption{Open-edition tiers of \textit{The Fungible} by Pak (2021). Collectors purchased cubes in varying quantities; the smart contract then minted unique NFTs whose visual composition reflected the number of cubes accumulated, turning fungible purchases into bespoke non-fungible artworks.}
    \label{fig:thefungible}
\end{figure}
\paragraph{Concept} \emph{The Fungible} (April 2021) was an ambitious collection that probed the definition of value in the context of digital art and fungibility. Created in partnership with Sotheby's, this project's very title is a play on the idea of fungible vs. non-fungible value: Pak offered an array of NFTs that collectively challenged the boundaries between unique art objects and interchangeable units. The concept centered on asking \emph{``what does value mean, and from where does it derive authority?''} \cite{Fungible2021}---a question that the collection explored by introducing novel mechanics (like token merging and dynamic editions) which blurred the line between a singular artwork and a set of exchangeable pieces. In essence, \emph{The Fungible} served as a critical examination of how scarcity and abundance intersect on the blockchain, turning the sale itself into an inquiry about why we value art: is it the content, the context of rarity, or the creative token logic behind it?

\paragraph{System Mechanism} \emph{The Fungible} deployed innovative smart-contract mechanics that made collecting an interactive process. Core to the system were the Open Edition ``Cubes,'' which were not traditional 1-of-n editions but a kind of modular token: buyers could purchase any number of identical cubes, and the contract would then deliver NFTs reflecting the quantities accumulated. For instance, if a collector purchased five cubes, they would receive an NFT depicting a cluster of five cubes; if they purchased fifty, they'd receive a single NFT showing fifty cubes---a graduated series that cleverly visualized ownership as a spectrum. This algorithmic distribution turned fungible purchases into bespoke non-fungible artworks. Additionally, Pak introduced dynamic one-of-one NFTs like \emph{The Switch}, coded such that its owner could decide to ``flip'' the artwork's state once at a moment of their choosing (an irreversible transformation built into the contract). Another piece, \emph{The Pixel}, consisted of a single gray pixel, emphasizing minimalism and the idea that even the smallest digital unit could hold enormous value in the right context. These mechanics---custom distribution, owner-triggered change, and extreme minimalism---showcased the versatility of smart contracts, engendering a set of artworks whose form and existence were defined by procedural rules and collector interaction, not just by static visuals.

\paragraph{Participatory Interaction} \emph{The Fungible}'s drop was designed as a multi-day, multi-modal event that actively involved the audience in creating and discovering value. Over a three-day period, collectors could participate in Open Edition sales where they bought ``fungible cubes'' in whatever quantity they wished during the sale period \cite{Fungible2021}, effectively letting each participant construct their own edition. Meanwhile, high-stakes auctions for one-of-a-kind pieces (\emph{The Switch} and \emph{The Pixel}) ran in parallel, drawing competitive bidding. Pak further engaged the community with puzzles and special rewards: for example, he gifted 30 unique NFTs to notable community ``builders'' and awarded four more NFTs (called ``Equilibrium'') to participants who met specific criteria like solving a puzzle or tweeting Pak's hashtag to the largest audience \cite{2021Burnart}. This layered approach meant that thousands of people joined at various levels---from casual buyers to dedicated players---making the sale itself feel like a collective happening. Every participant, whether buying a single cube or vying in an auction, became part of the narrative fabric that \emph{The Fungible} wove around the notion of value.

\paragraph{Artist's Control} Once \emph{The Fungible} launched, Pak largely let the programmed systems and collectors drive the outcomes. The rules for minting cubes and the parameters for pieces like \emph{The Switch} were all predetermined, meaning that after initiating the sale, Pak did not manually influence how many cubes were sold or how the artworks evolved. In fact, \emph{The Switch} exemplified a deliberate transfer of artistic agency: the moment of its transformation was controlled by the collector, not the artist \cite{Fungible2021}. Similarly, the open edition ran on its own schedule---Pak set the prices and time windows for each cube sale, but could not alter the open-edition once it was live. By structuring the project this way, Pak's role became one of orchestrator rather than micromanager. He created the conditions and constraints, then ceded control to the market and the code, a choice consistent with the work's exploration of decentralized value creation. After the sale, the artworks existed independently on the blockchain (e.g., \emph{The Pixel} perpetually remaining just one pixel owned by its auction winner), requiring no further intervention from the artist.

\paragraph{Participants \& Collective Emerging Behavior} \emph{The Fungible} attracted an unprecedented scale of participation for an NFT art drop at the time. Over 3,000 individuals took part in the open edition cube sale alone, resulting in a total of 23,598 cubes sold across three short selling periods and 6,165 unique NFTs minted from the open edition process. The event's gamified structure (with leaderboards for top cube buyers, surprise rewards, and the allure of auctions) encouraged a spirited community response. Buyers strategized over how many cubes to purchase to attain higher-tier cube NFTs, while others collaborated or competed for the special rewards. Sotheby's reported record-breaking engagement on Nifty Gateway during the sale, including the most bids ever placed on an NFT auction item up to that point. This collective excitement transcended typical art-world transactions; it felt more akin to a massive multiplayer event in which the audience's size and enthusiasm were integral to the work's impact. In the aftermath, collectors and observers alike debated whether the frenzy and competitive ``game'' of \emph{The Fungible} were integral to its value, or if those incentives would fade, directly engaging with the very question the artwork posed about how value is sustained in digital art.

\paragraph{Poetic Meaning-Making} Beyond its clever mechanics and record-breaking sales, \emph{The Fungible} resonated as a poetic statement on the fluid nature of value in the digital age. By literalizing the idea of fungibility---letting collectors accumulate and assemble value units (cubes)---Pak drew attention to the transactional element of art without forfeiting aesthetic intrigue. The collection as a whole becomes an allegory: the cubes symbolize the building blocks of value, and their conversion into unique artworks symbolizes the alchemy by which markets and perception turn the mundane into the precious. The open question \emph{The Fungible} leaves us with is whether art's value is intrinsic or constructed: when a single pixel sells for \$1.36 million and thousands of identical cubes become rare sculptures through a smart contract, one is compelled to acknowledge the role of consensus, context, and collective belief in creating worth. In the broader digital art discourse, \emph{The Fungible} is seen as a work that married participatory art with commentary on economics, ultimately reminding us in a poignant way that even on the blockchain, value is a story we all collectively tell.
\paragraph{Key Statistics} \textit{The Fungible} drop took place 12--14 April 2021 on Nifty Gateway, generating approximately \$16.8 million in sales over three days. The open edition ``Cubes'' were sold in timed windows at \$500, \$1,000, and \$1,500 price points on successive days, with a total of 23,598 cubes purchased. These merged into 6,165 NFTs, distributed to 3,080 unique buyers (reflecting many bought multiple cubes). Sotheby's auctioned \textit{The Pixel} for \$1.36 million and \textit{The Switch} for \$1.44 million, each becoming one of the earliest NFT artworks to surpass \$1 million. The special one-of-one \textit{The Cube} was awarded to the top cube buyer, and 100 editions of \textit{Complexity} were awarded to the next 100 top buyers. In total, over 3,080 unique buyers participated in the primary sale. At the time, it set a record for the largest NFT sale by a single artist on a platform, until Pak's own \textit{Merge} later that year.

\paragraph{Annotation} \emph{The Fungible} is the portfolio's strongest early case for \textbf{C5} (economy-driven engagement): the Sotheby's partnership, tiered pricing, leaderboard competitions, and puzzle rewards made the market event indistinguishable from the artwork. It equally demonstrates \textbf{C1} (system-centric composition), as the algorithmic conversion of fungible cubes into unique NFTs---whereby participants could purchase interchangeable units and watch an algorithm transmute them into bespoke non-fungible artworks---embodies the shift from object to process. Crucially, the work plays with the very boundary between fungible and non-fungible: collectors could combine non-fungible elements back into fungible form, a kind of ``coin play'' that literalizes the conceptual slippage between currency and art. \textbf{C3} (distributed agency) is present but structured---participants co-determine the outcome through purchasing decisions, yet within tightly scripted rules, distinguishing this from the more open-ended authorship of Lost Poets or Censored. What elevates \emph{The Fungible} beyond a single project, however, is its role as the foundational experiment in composability for Pak's subsequent practice. The market here provides an interaction space in which users can play, combine elements, and test composability, and the mechanics piloted in \emph{The Fungible} directly laid the groundwork for the token-burning logic of burn.art, the multi-stage transformations of Lost Poets, and the self-merging dynamics of Merge. In this respect, \emph{The Fungible} constitutes an early-stage laboratory for \textbf{C7} (interoperability and composability), leading users toward open-ended authorship and establishing the paradigm that Pak would elaborate across his entire oeuvre.

\paragraph{Information}

\begin{itemize}
\item 
Launch Website: \url{https://www.niftygateway.com/collections/paksothebysauction/}
\item 
Release Date: Apr 12, 2021
\item 
OpenSea: \url{https://opensea.io/collection/the-fungible-by-pak}
\item 
Contract Address: \\0xc7cc3e8c6b69dc272ccf64cbff4b7503cbf7c1c5 (ERC721)
\end{itemize}

\subsection{burn.art / \$ASH (2021)} 

\begin{figure}[ht]
    \centering
    \includegraphics[width=0.9\linewidth]{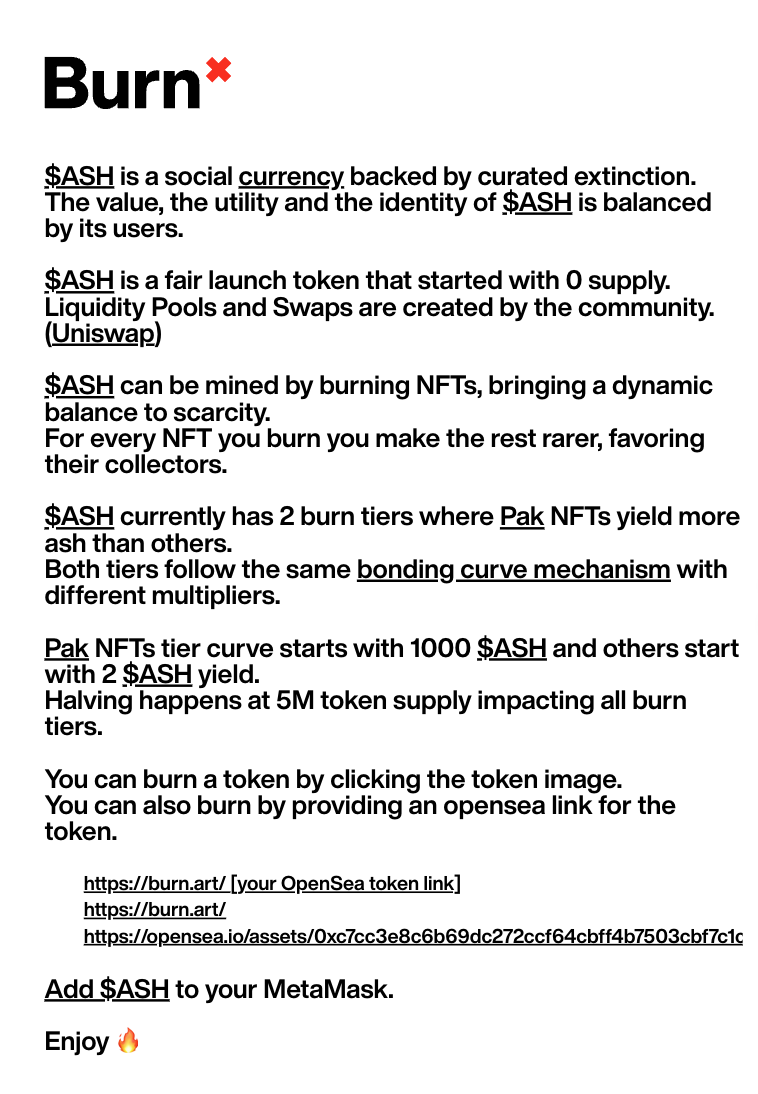}
    \Description{A text-based screenshot of the burn.art website explaining the \$ASH token system. The page describes \$ASH as a social currency backed by curated extinction, explains that it can be mined by burning NFTs, details a two-tier system where Pak NFTs yield more \$ASH, mentions a bonding mechanism with different multipliers, a halving event at 1,000 \$ASH and another at 5 million token supply, and provides instructions for adding \$ASH to MetaMask.}
    \caption{The burn.art landing page (2021), outlining the \$ASH token economy. The system formalizes ``creation through destruction'': collectors permanently burn NFTs to receive \$ASH tokens, establishing a cyclical protocol where artistic value is destroyed, transmuted, and regenerated.}
    \label{fig:burnart}
\end{figure}

\paragraph{Concept} \emph{burn.art} (and its native token \$ASH) is Pak's poetic framework for \emph{``creation by destruction,''} an artwork-platform that turns the act of burning NFTs into an artistic medium in itself. Launched in 2021 after \emph{The Fungible}, it proposed a cyclical ecosystem: collectors are invited to destroy their existing NFT artworks (irreversibly sending them to a burn address), and in return they receive a new fungible token called ASH---essentially the ``ashes'' of the burned art \cite{2021Burnart}. These ashes (\$ASH) can then be spent on new art pieces by Pak, or even other creators, thereby continuing the cycle of creative destruction. The conceptual core is a commentary on value and permanence: by explicitly linking loss and creation, Pak challenges the notion of digital art as eternally replicable or static. Instead, \emph{burn.art} frames destruction as a generative act, asking us to consider what we value more---the art we had, or the new possibilities unlocked by its sacrifice.

\paragraph{System Mechanism} The underlying mechanism of \emph{burn.art} is an Ethereum smart contract that accepts NFTs and issues ERC-20 ASH tokens in return. Technically, when a user initiates a burn, the NFT is transferred to an inaccessible null address (removing it from circulation forever), while the contract credits the user with a certain amount of ASH according to a predefined schedule. The supply of ASH is thus dynamically generated by destruction: every token in circulation is proof that some piece of digital art was sacrificed. This introduces a self-balancing scarcity model---the more art is burned, the rarer the remaining artworks become, favoring their collectors by increasing relative scarcity. Pak designed \$ASH's tokenomics such that all his future drops could interact with this system, for example by accepting only \$ASH as payment for certain new works or giving exclusive access to ASH holders. This created an evolving feedback loop: the dynamics of burning and creating continually affect each other, crafting a mini-economy where collectors' decisions to destroy or hold collectively shape the progression of the art ecosystem.

\paragraph{Participatory Interaction} Using \emph{burn.art} is itself a performative interaction: collectors must actively choose to irreversibly destroy one of their NFTs in order to partake. Upon connecting their crypto wallet to the burn.art interface, a user can select an NFT they own and send it to the burn contract; the system then ``mines'' (mints) a corresponding amount of ASH tokens as a reward. The amount of ASH received depends on the category of the burned NFT---Pak implemented a whitelist with multipliers, meaning more significant or scarce works yield more \$ASH. This process engages users on a psychological level: burning an artwork, especially a valuable one, is a moment of suspense and conviction, effectively making the user a co-creator of the conceptual piece. Communities sprung up around sharing burning experiences and strategies (e.g., which NFTs to burn for optimal \$ASH yield), turning what could have been a solitary act into a communal ritual. By requiring collectors to ``prove'' their dedication (through destruction) to gain access to new creations, \emph{burn.art} ensured that participation was not passive but deeply intentional and symbolic.

\paragraph{Artist's Control} While \emph{burn.art} introduced a new level of community agency (the choice of what and when to burn lies with the users), Pak maintained strategic control over the ecosystem's parameters. He defined which NFTs are burnable for ASH and their relative yields via the whitelist---essentially curating the value of destruction. Furthermore, as the creator of the ASH token, Pak could influence its use by deciding which new artworks require ASH for purchase, thereby indirectly guiding demand for burning. That said, once the contracts were deployed, the day-to-day operation of the system is autonomous: any eligible NFT can be burned permissionlessly, and \$ASH is dispensed according to code, not at Pak's discretion. In keeping with decentralization, the value of ASH and the decision of what to sacrifice for it were left to the open market and individual collectors. Pak's control was thus front-loaded---in the design and rules of the game---but he stood back as participants carried out the dramatic act of creation-through-destruction, letting the social experiment unfold organically.

\paragraph{Collective Emergent Behavior} \emph{burn.art} quickly garnered a devoted following of Pak's collectors and intrigued onlookers, forming a sub-community fixated on the burn-to-mint ritual. Hundreds of NFTs were burned in the initial weeks, from common items to high-value artworks, as collectors sought to accumulate \$ASH either for status or to use in Pak's future drops. This created a shared ethos among participants: a mix of fanaticism and camaraderie emerged, where those who burned valuable assets were celebrated (or playfully lamented) for their commitment to the art. Discussions on forums and social media centered on what NFTs people were willing to sacrifice; the ``burn culture'' introduced by Pak made some collectors view their holdings in a new light (knowing any Pak piece, if desired, could be converted to \$ASH, a kind of artistic afterlife for stagnant assets). Over time, the collective action of these participants literally shaped Pak's art market: as more tokens were destroyed, the remaining ones grew rarer, and owning \$ASH became a badge of participation in Pak's evolving narrative. This self-selected community, willing to destroy to create, exemplified a new kind of collector ethos unique to the protocol---one that values conceptual engagement over simply accumulating objects.

\paragraph{Poetic Meaning-Making} \emph{burn.art} and \$ASH elevate an almost mythic cycle of death and rebirth within the digital realm, making it one of Pak's most overtly metaphorical works. The project transforms the destructive act of blockchain burning into a meaningful ritual, suggesting that even in a space of infinite reproducibility, sacrifice can carry weight and give rise to new beauty. The poetry here lies in how absence is turned into presence: each \$ASH token is a memorial of a destroyed artwork and simultaneously the seed of a new creation. In a world where digital files ostensibly last forever, Pak introduced impermanence and choice as artistic elements---echoing the destruction of physical art (recalling Gustav Metzger's ``auto-destructive art'' or Jean Tinguely's self-destroying sculptures) but with a constructive twist. The broader impact of \emph{burn.art} is its commentary on value: it forces the community to question what gives an artwork value---its continued existence, or the legacy it leaves through transformation. By making his audience perform a collective dance of loss and renewal, Pak illustrated that in digital ecosystems, as in nature, endings can be beginnings. This piece thus bridges conceptual art and social experiment, leaving a lasting impression about the responsibilities and powers that protocols confer upon both artists and collectors.
\paragraph{Key Statistics} \textit{burn.art} launched in May 2021. Within the first 48 hours, users burned over 1,000 NFTs to generate \$ASH. By February 2022, more than 20,000 NFTs had been burned via the platform, including high-profile pieces by Pak and other artists, yielding over 4 million \$ASH tokens. \$ASH reached a peak market capitalization above \$100 million during the 2021 NFT boom, reflecting the community's speculative interest. Pak's first \$ASH-only NFT drop, ``Chapter One---Carbon,'' sold out in minutes, consuming a substantial portion of the token supply. The initial reward rate was 1000 \$ASH for Pak NFTs and 2 \$ASH for others, with an automatic halving mechanism that reduces rewards at each supply milestone. As of 2025, over 30 million \$ASH have been minted through successive burns, and the token is used by a handful of other crypto artists who accept \$ASH for their work, extending Pak's poetic economy of art and \$ASH.

\paragraph{Annotation} \emph{burn.art} is the pivotal work for \textbf{C7} (interoperability and composability): it introduces \$ASH as a cross-project fungible token that bridges The Fungible, Lost Poets, Merge, and future drops, transforming Pak's oeuvre from a series of discrete works into a unified economic ecosystem---the center of the Pak universe. It is also a powerful case for \textbf{C6} (poetic message embedding), as the ritual of destruction-as-creation literalizes themes of sacrifice, renewal, and the cyclical nature of value. \textbf{C2} (autonomous governance) is strong---the burn contract operates permissionlessly---yet Pak retains unusual ongoing influence through an ultimate DAO behind the contract that can tune burning rates for each series, granting dynamic ability to control floor prices across the entire collection. This mechanism functions as an economic adjustment machine: if the floor price of any Pak series drops below the potential \$ASH yield from burning, buyers have a rational incentive to purchase and burn those NFTs into \$ASH, thereby automatically establishing a guaranteed floor price. The structure operates, in effect, like a Federal Reserve for Pak's own art---a brilliantly engineered economic mechanism that guarantees basic value while preserving market freedom. This hybrid of autonomy and curatorial control complicates any simple narrative of decentralization in protocol art, making burn.art the only project where the artist actively steers post-launch demand through parametric governance rather than direct intervention. In its synthesis of permissionless burning, cross-project interoperability, and dynamic economic tuning, \emph{burn.art} is arguably the most ingenious infrastructure-level artwork in the portfolio: an incredibly smart mechanism that turns the entire ecosystem into a self-regulating economy of creative destruction.

\paragraph{Information}
\begin{itemize}
\item 
Launch Website: \url{https://burn.art/}
\item 
Release Date: May 16, 2021
\item 
Contract Address: \\0x64d91f12ece7362f91a6f8e7940cd55f05060b92 (ERC20)
\end{itemize}

\subsection{Lost Poets}

\begin{figure}[ht]
    \centering
    \includegraphics[width=0.9\linewidth]{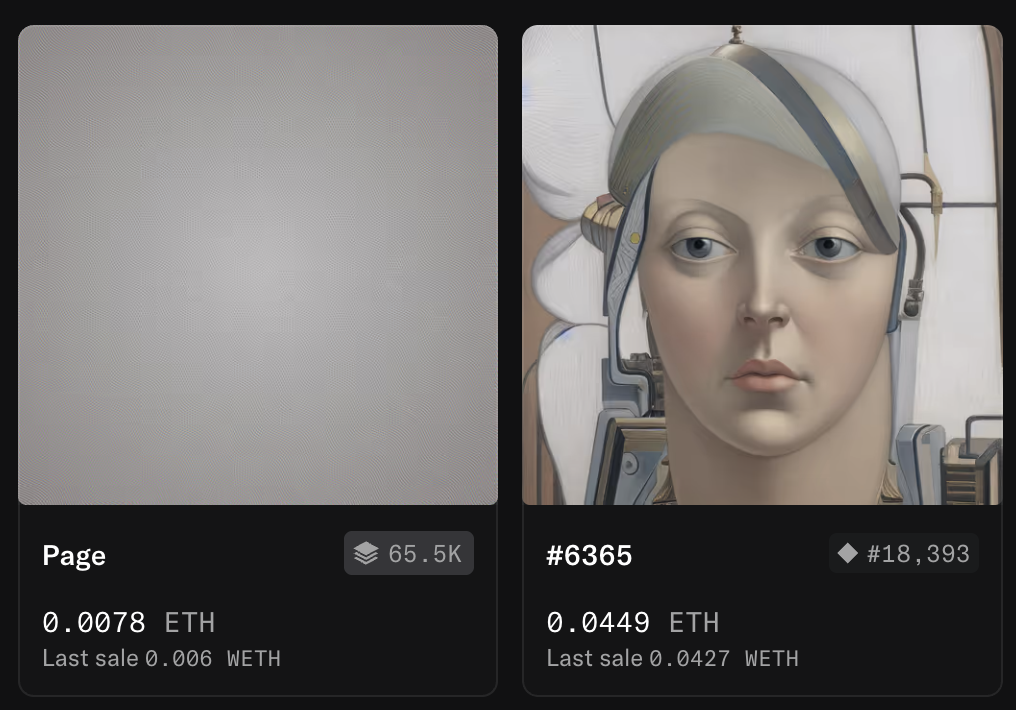}
    \Description{Two NFT listings side by side on a marketplace. Left: a blank gray square labeled ``Page'' priced at 0.0078 ETH, representing an unrevealed token. Right: an AI-generated portrait labeled number 6365 priced at 0.0449 ETH, showing a Renaissance-style female face with mechanical elements, representing a revealed Poet.}
    \caption{Lost Poets: before and after the reveal. Left: an unrevealed Page token, a blank placeholder. Right: a revealed Poet (number 6365), an AI-generated portrait unlocked by burning the Page---illustrating the irreversible state transition at the heart of the project's multi-act protocol.}
    \label{fig:lost-poets}
\end{figure}

\begin{figure*}[ht]
    \centering
    \includegraphics[width=0.9\linewidth]{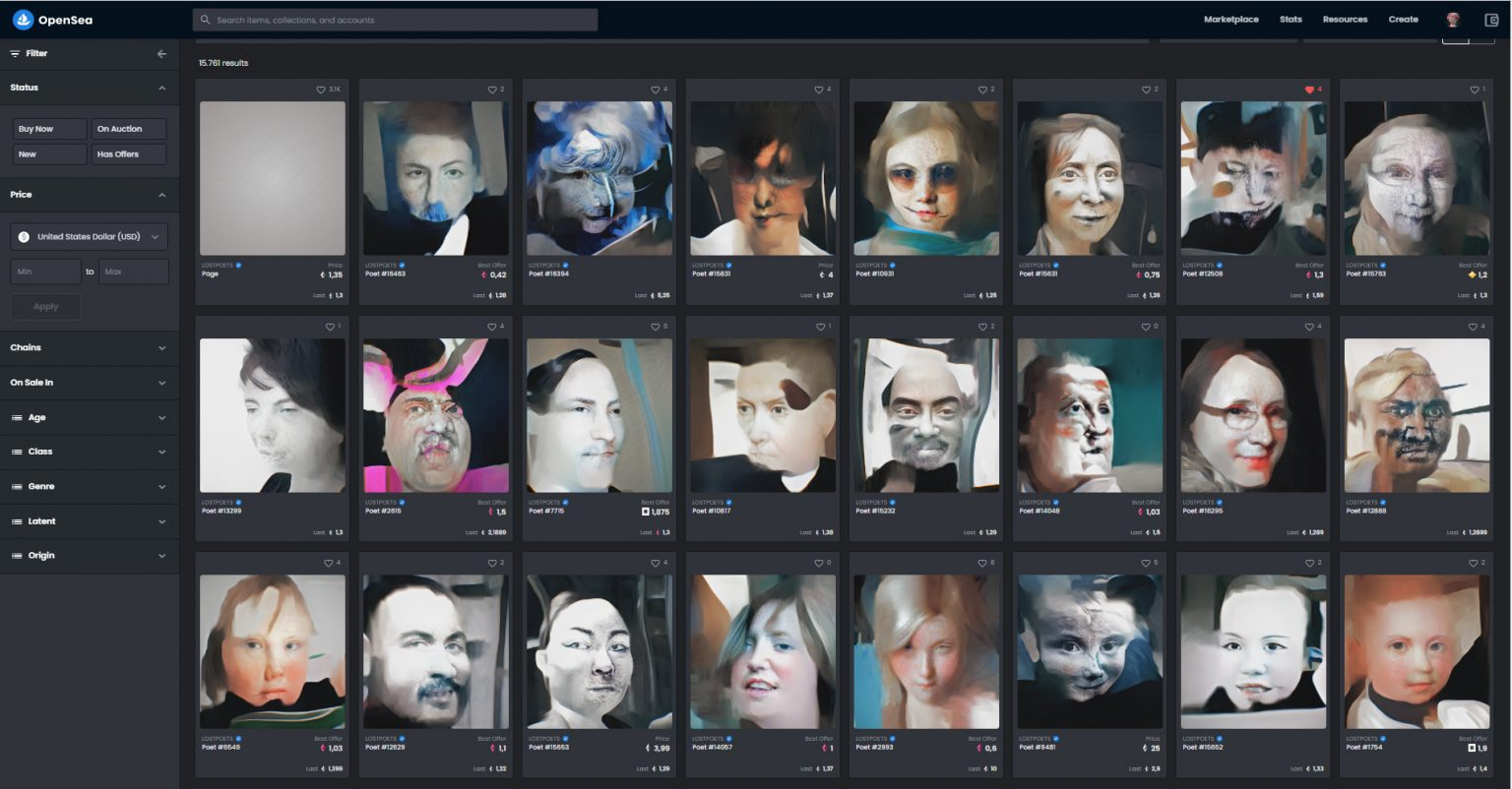}
    \Description{A wide screenshot of an OpenSea-style gallery showing approximately thirty revealed Lost Poets NFTs in a grid. Each displays a unique AI-generated portrait in diverse styles---some photorealistic, some painterly, some surreal---depicting faces of varying ages, ethnicities, and visual treatments. A sidebar on the left shows filtering options for collection attributes.}
    \caption{Gallery view of revealed Lost Poets on the secondary market. Each portrait was procedurally generated with unique visual traits; collectors further individualized their Poets by naming them and ``feeding'' additional Page tokens to unlock poetic word combinations, layering human authorship onto algorithmic creation.}
    \label{fig:Lost-Poets2}
\end{figure*}

\paragraph{Concept} \emph{Lost Poets} (September 2021) is a large-scale NFT art project that merges generative art, narrative puzzle, and strategic game into one poetic whole. At its core, the project consists of 65,536 unique AI-generated portrait NFTs (``Poets'') originating from 1,024 rarer ``Origin'' Poets \cite{Anderson2022Lost}, all imbued with an aura of mystery and antiquity. The concept draws inspiration from Borges's \emph{Library of Babel} and the idea of rediscovering lost creative souls: Pak envisioned these Poets as ancient voices ``not found'' until the community brings them to light. The project was unveiled as a multi-act saga---a kind of unfolding story where the NFTs would evolve over time---thereby exploring themes of memory (lost knowledge rediscovered), evolution (the NFTs reveal attributes in phases), and collective intelligence. \emph{Lost Poets} is fundamentally about the interplay between algorithmic creation (the AI art and smart-contract logic) and human participation in meaning-making: it asks what happens when thousands of collectors jointly play a literary game with art, time, and chance.

\paragraph{System Mechanism} The \emph{Lost Poets} smart contracts orchestrated an intricate multi-stage evolution of the NFTs. Initially, 65,536 Page tokens were minted (with Pak reserving 1,024 for the special Origin Poets). These ERC-1155 tokens had built-in utility: a holder could ``burn'' a Page token to transform it into a Poet token once Act II commenced. The transformation was irreversible, effectively migrating the NFT from one state (Page) to another (Poet) and revealing the AI-generated portrait and some initial attributes. Crucially, the contract did not reveal all attributes of a Poet at once; instead, additional traits (like words the Poet could ``speak'') unlocked gradually over Act II and Act III, either automatically or through user actions. Feeding a Page to a Poet (another burn action) was a method to update the Poet's metadata---the contract would record the new name given by the user and randomly assign a set of words to that Poet, altering its uniqueness. All these state changes were governed by on-chain logic, ensuring that no two Poets ended up the same despite coming from identical Pages. The system also tracked leaderboards and triggers for awarding the Origin Poets to top collectors at the end of Act I. Finally, a time-lock dynamic was present: at the conclusion of the one-year project timeline, the contract ``locked in'' the final state of all Poets (Act IV, aptly titled ``The Twist'') and dispensed promised \$ASH rewards to participants, marking the end of the journey. This layered smart contract design allowed \emph{Lost Poets} to behave almost like an autonomous game engine running on Ethereum, driving the content and rarity of the NFTs in response to player inputs and time. 

\paragraph{Participatory Interaction} Engagement with \emph{Lost Poets} was deeply interactive and unfolded in distinct phases (Acts I--IV). It began with the distribution of ``Pages''---blank NFTs that served as tickets to the game \cite{Lostpoets}. Early on, Pak rewarded loyal fans by airdropping Pages to wallets holding at least 25 \$ASH tokens and opened a public sale where others could purchase Pages (priced at 0.32 ETH each). Once armed with Pages, participants faced choices in Act II (``The Reveal''): they could burn a Page token to summon its Poet (revealing a unique AI-crafted portrait), or they could even burn Pages for more \$ASH instead, adding a twist of sacrifice to the game. In Act III (``The Explorer''), collectors could feed extra Pages to their Poets, each feed allowing them to rename the Poet and gifting the character new words (2--4 random words per Page) which enriched its traits and lore. Throughout these interactions, Pak maintained a live leaderboard; the top 100 collectors who amassed the most Pages received additional rare Origin Poets as rewards, incentivizing strategic play and competition. The project was designed to span 365 days, during which participants continuously engaged---naming their Poets, deciphering clues, trading Pages and Poets, and anticipating ``The Twist'' of Act IV. This sustained, game-like interaction meant that collecting \emph{Lost Poets} was not a one-off transaction but a prolonged creative exercise that bound the community together.

\paragraph{Artist's Control} Pak's role in \emph{Lost Poets} was akin to a dungeon master setting the rules of a game, then letting the players roam. Pak exercised significant control in the design phase: determining the number of tokens, the pacing of the acts, the mechanics for conversion and feeding, and even withholding the 1,024 Origin Poets to distribute as rewards or for future surprises. However, once the project commenced, the progression was largely automated and community-driven. Pak did not alter the contract rules mid-course; the code itself dictated when new attributes would appear and how each action translated into an outcome. The element of the unknown (``The Twist'' of Act IV) was built-in---Pak had likely pre-planned an ending but kept it secret, triggering it through the programmed schedule or an on-chain call at the appropriate time. Throughout the year-long evolution, Pak's direct intervention was minimal, aside from providing occasional hints or narrative flavor via social media to enrich the lore. In essence, after launching \emph{Lost Poets}, Pak relinquished control to the protocol and the participants, allowing the artwork to self-evolve within the boundaries Pak had coded. This balance---tight authorial control over structure, but freedom for users within it---underscores Pak's commitment to exploring protocol as an art form.

\paragraph{Collective Emergent Behavior} \emph{Lost Poets} attracted a massive and engaged community, as evidenced by the entire supply of 65,536 Pages selling out in just 2 hours (raising about \$70 million) and thousands of collectors joining the fray. The project's gamified nature led to rich emergent behaviors: collectors formed online groups to decode hidden clues in Poet attributes and to strategize the best use of Pages. A vibrant secondary market developed where Pages and Poets were traded, with some speculators hoarding Pages early in hopes of obtaining more Origins or leveraging them in later acts. The competition for the top 100 collector spots was intense, leading certain individuals to accumulate enormous quantities of Pages (and become known figures in the community for it). Meanwhile, other participants took a more curatorial approach, carefully naming their Poets and sharing the whimsical or profound word combinations their Poets ``spoke'' after feeding---effectively collaboratively writing a decentralized poem through their NFTs. Each phase change (Act II's reveal, Act III's feeding, etc.) was accompanied by collective excitement on forums and Discord, as people shared discoveries and theories about what Act IV, ``The Twist,'' might entail. By the end of the 365 days, \emph{Lost Poets} had fostered not just a market but a participatory culture---blending competition, storytelling, and collaboration---arguably one of the most sustained communal engagements for an NFT project of its era.

\paragraph{Poetic Meaning-Making} The poetry of \emph{Lost Poets} emerges through its synthesis of technology and storytelling. On one level, it is a meditation on authorship: each collector who names their Poet and contributes words becomes a co-author, blurring the line between artist and audience in a manner reminiscent of exquisite corpse or collaborative poetry. The project also poignantly plays with the idea of ``lost'' voices found---the algorithm conjured tens of thousands of unique faces and fragments of text, like ghosts of poets past, and gave the community the power to nourish these ghosts with new words. Over the course of the acts, participants experienced themes of discovery, transformation, and ephemerality (as unused Pages dwindled and choices had to be made). In the final Epilogue, when all Poets reached their ultimate form and the last secrets were unveiled, the community could look back on a journey that it had collectively authored. Pak managed to reveal that NFTs need not be static collectibles; they can be alive with narrative potential and participant-driven change. \emph{Lost Poets} stands as a testament to the idea that protocol-based art can achieve a form of literary and artistic richness---it turned a blockchain ledger into the stage for a year-long participatory saga about remembering and reinventing creative voices. The broader impact is an expanded notion of what digital art can be: not just an image or an object, but a living story that engages its audience in poetic meaning-making.

\paragraph{Key Statistics}  \textit{Lost Poets} launched with 65,536 Page NFTs at 0.32 ETH each, selling out and raising roughly \$70 million in its initial sale. 1,024 Origin Poets were distributed (277 to top 100 collectors, 730 via daily random drops, 17 reserved). Phase II saw thousands of Pages burned: within the first week, over 50,000 Pages were converted into Poet NFTs, leaving near 15,000 Pages for the final phase. Each Poet NFT is programmatically one of 1,024 ``families'' and has 256 trait parameters, making each uniquely identifiable. As of the end of Phase III, about 15,000 Poets received names and poems (meaning that many Pages were sacrificed to finalize them), while a number of Poets remained ``silent'' because their owners chose not to burn pages. The most popular names taken include historical luminaries (``Shakespeare'' was named within minutes of launch). Secondary market activity was robust: Origin Poets traded at a premium (some over 5 ETH each) due to their limited supply and role as progenitors. By project's end, the community had collectively written tens of thousands of individual one-line poems, effectively creating one of the largest collaborative literary works in NFT form.

\paragraph{Annotation} \emph{Lost Poets} is the richest case for \textbf{C3} (distributed agency) and \textbf{C4} (temporal dynamism): its multi-act structure unfolds over a full year in a genuinely performative and crowdsourced process, and collectors become genuine co-authors by naming their Poets and choosing whether to transform, hold, or burn their tokens. The temporal dynamic is essential: each phase---from the initial Page distribution through the Reveal, the Explorer phase, and the final Twist---constitutes a distinct participatory epoch, and the work's meaning accrues through the cumulative decisions of thousands of participants across these stages. It also deeply exemplifies \textbf{C6} (poetic message embedding), with the Borges-inspired narrative framing the naming ritual as a meditation on language and memory. \textbf{C7} (composability) is significant: the project bridges into the \$ASH ecosystem, utilizing the burn mechanism introduced in burn.art---participants can burn Poets for \$ASH, and \$ASH holders received airdropped Pages, creating a feedback loop between projects. What makes \emph{Lost Poets} distinctive is the depth of its branching decision space---at each stage, every participant faces a choice to transfer, hold, or burn, and each holder's path through the Acts yields a different outcome, producing an emergent diversity that exceeds what any single-mechanism work (like X or Merge) can generate. The project demonstrates that everyone can participate in the unfolding of protocol art, and that the temporally extended, multi-stage structure is essential for cultivating the kind of sustained collective authorship that transforms a smart contract into a living literary work.

\paragraph{Information}

\begin{itemize}
\item 
Launch Website: \url{https://lostpoets.xyz/}
\item 
Release Date: Sep, 2021
\item 
OpenSea: \url{https://opensea.io/collection/lostpoets}
\item 
Contract Address: 
\\0xa7206d878c5c3871826dfdb42191c49b1d11f466 (ERC1155 for Page)
\\0x4b3406a41399c7fd2ba65cbc93697ad9e7ea61e5 (ERC721)
\end{itemize}

\subsection{Hate / Invisible Mechanisms (2021)}

\begin{figure}[ht]
    \centering
    \includegraphics[width=0.9\linewidth]{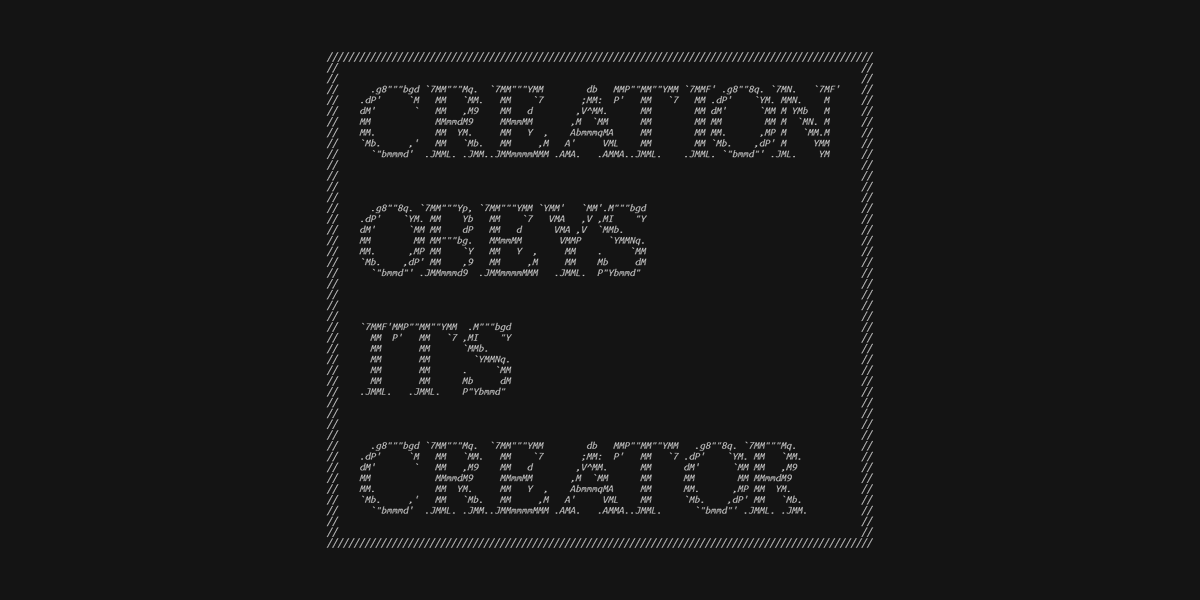}
    \Description{Large ASCII-art text rendered in monospace characters on a dark background spelling out ``CREATION OBEYS ITS CREATOR'' in three lines. Each letter is composed of smaller text characters that appear to be fragments of smart contract source code, visually fusing the philosophical statement with the underlying program code.}
    \caption{``CREATION OBEYS ITS CREATOR''---a statement embedded as ASCII art in the source code of the \textit{Hate} NFT smart contract, part of Pak's \textit{Invisible Mechanisms} (2021). The phrase encapsulates the \textit{Move} protocol's radical inversion of NFT ownership: the creator retains administrative power to lock, transfer, or reclaim any token at will.}
    \label{fig:invisible}
\end{figure}

\paragraph{Concept} \emph{Invisible Mechanism} are intertwined artworks that together form a provocative commentary on control, ownership, and the relationship between creator and audience (even the antagonistic part of the audience). In late 2021, Pak executed \emph{Hate} as a conceptual performance: Pak offered special NFTs titled ``Hate'' to thirty of Pak's harshest critics on social media---effectively gifting tokens to those who had publicly disparaged Pak \cite{Pak2021Haters}. The twist was that these \emph{Hate} NFTs were programmed to be immutable and non-transferable by the recipients, symbolizing the idea that their negativity had earned them a token they could not get rid of. This gesture served as both a prank and a statement, turning the act of hating Pak into an involuntary participation in Pak's art. Underpinning \emph{Hate} was Pak's new smart contract prototype called \emph{Move}: an ``invisible mechanism'' that grants the creator the power to transfer any token from any wallet within that contract to another address at will, irrespective of holder consent. \emph{Move} embodies the mantra ``CREATION OBEYS ITS CREATOR'', flipping the usual decentralization narrative to assert total artistic control over distribution. Together, \emph{Hate} (the scenario) and \emph{Move} (the technology) form a conceptual piece that asks pointed questions: What if an artist could reclaim authority over an artwork after it's been distributed? Can an artwork be a form of dialogue---or retaliation---between artist and critic? By engaging directly with haters and by unveiling a contract that subverts NFT norms, Pak's concept challenged the community to reflect on power dynamics and the presumed inviolability of ownership in the crypto space.

\paragraph{System Mechanism} The \emph{Hate} NFTs were engineered using the \emph{Move} contract architecture, which fundamentally alters standard ERC-721 behavior \cite{2021PakReveals}. Normally, once an NFT is in a wallet, only the owner can transfer it; Pak's custom contract introduced admin functions that allowed Pak (as the contract owner) to lock tokens and later move them at will. Upon minting the 30 \emph{Hate} tokens, they were immediately ``locked'' to their respective recipient addresses---hence any transfer or sale initiated by those holders would automatically fail, as the contract overrode such actions. The NFTs had identical metadata (depicted simply by a heart symbol), emphasizing that their differentiating feature was the wallet they resided in, not visual content. After letting the scenario play out for about 24 hours, Pak utilized \emph{Move} to demonstrate its power: Pak, as admin, transferred all \emph{Hate} tokens out of the haters' wallets back into Pak's own, even updating their metadata before doing so. This action conclusively illustrated the dynamic---the artist could literally reclaim the tokens at will, something impossible under standard token contracts. The \emph{Move} contract thus created a dynamic where token ownership was provisional, entirely subject to the creator's whims. Such a mechanism is invisible until exercised, earning its name, and it turned the typical trust model of NFTs on its head: here, the code enforced a hierarchy where the art always ultimately ``belonged'' to its maker.

\paragraph{Participatory Interaction} \emph{Hate} inverted typical participation---it was the artist who ``participated'' in the lives of certain audience members by depositing NFTs into their wallets uninvited. Pak solicited volunteers by asking Pak's followers to prove they had posted genuine hate toward Pak (via old tweets). The ``winners'' of this dubious honor were then airdropped the \emph{Hate} NFTs; about 30 individuals received one. Their engagement was largely one of surprise and frustration: many attempted to trade or remove the tokens, only to discover every transfer attempt was automatically rejected by the contract. In essence, their interaction was to experience powerlessness---an unusual, perhaps uncomfortable form of audience participation where the usual agency of a collector was stripped away. As for \emph{Move}, the wider community's participation came through observation and discussion rather than direct use (since \emph{Move} is a contract design, not a direct-to-consumer app). Pak publicly shared the \emph{Move} contract and its implications for all to inspect, inviting creators and collectors to contemplate this new paradigm. In doing so, the community collectively engaged in a discourse about consent and control in NFTs. In summary, \emph{Hate/Move} turned the tables on engagement: instead of people actively opting into the art, the art opted into their lives, thereby making the audience's reactions and discussions (especially the haters' public bemusement) an integral part of the piece.

\paragraph{Artist's Control} By its very nature, \emph{Hate/Move} was an exercise in maximizing artist control. Pak retained absolute authority over the \emph{Hate} tokens at all times---Pak was the only one who could mint them, the only one who could move or burn them, and even the only one who could change their content. This project was a deliberate outlier compared to Pak's other works: instead of surrendering control to autonomous rules or the community, Pak crafted a scenario where Pak's hand remained firmly on the lever from start to finish. In fact, the project can be seen as a manifesto of control: \emph{Move} was Pak asserting that as an artist-programmer, Pak could imbue a token with a piece of Pak's will, overriding the default freedoms typically given to collectors. After the initial performative phase, Pak ended the experiment on Pak's own terms by retrieving the tokens and effectively erasing them from the participants' wallets. In doing so, Pak underlined the central statement: these creations were never truly ``theirs''---they were on loan at the creator's mercy. Such extreme control is generally anathema to the decentralization ethos, which is precisely why \emph{Hate/Move} is so incisive; it used exaggerated artist control as an artistic device itself.

\paragraph{Collective Emergent Behavior} The immediate participants---the recipients of \emph{Hate}---reacted with a mix of confusion, amusement, and annoyance. Some joked about selling their entire wallet (since they couldn't sell the token itself), highlighting the absurd lengths one would need to go to bypass the restriction. This small group of 30 became inadvertent performers in Pak's drama, their attempts and public commentary forming a crucial part of the narrative that others followed. Meanwhile, the broader NFT community watched the spectacle with fascination. On social media and blogs, a flurry of debate ensued about the ethics and implications of \emph{Move}: some collectors expressed relief that such a mechanism was not widespread, while some creators were intrigued by the power it demonstrated. The collective behavior here was largely discursive---\emph{Hate/Move} spurred conversations about the nature of ownership and the extent of smart contract programmability. People questioned how this contrasts with the spirit of self-sovereignty that NFTs usually promise, and whether it was ``fair'' or simply a clever conceptual stunt. In the end, once Pak reclaimed the \emph{Hate} NFTs, the haters were free of their burdens, perhaps a bit wiser to the power of code in digital art. The community that observed gained a nuanced understanding: emerging from it was a collective acknowledgement that decentralization in art is not absolute, and that the artist's intent coded into a contract can drastically shape user experience. This dialogue and reflection were exactly the emergent cultural outcomes Pak likely sought, elevating the work from a mere prank to a significant case study in the NFT space.

\paragraph{Poetic Meaning-Making} In retrospect, \emph{Hate/Move} reads like a sharp, if mischievous, parable within the NFT art narrative. It poetically frames the tension between artist and audience: the \emph{Hate} tokens were as much a mirror to the recipients (reflecting their hostility back as an unusable ``gift'') as they were a canvas for Pak's statement about artistic sovereignty. There's dark humor and irony in the idea that hatred towards an artist could be alchemized into an artwork that essentially traps that hate---a modern digital twist on holding up a mirror to one's critics. The \emph{Move} mechanism adds an extra layer of meaning: it challenged the community to realize that the liberties they take for granted in decentralized art can be tweaked or overturned by a clever creator. This provocation was unsettling to some and thrilling to others, serving as what one might call a necessary thought experiment in a hype-driven field. By co-opting Pak's detractors into unwilling collaborators, Pak turned negativity into a generative component of Pak's practice, raising questions about consent, power, and the very definition of ownership. In the broader context of digital art discourse, \emph{Hate/Move} is significant for revealing that the blockchain is not inherently liberating---its effect depends on how it's used. Pak's broader impact here is to remind both creators and collectors that the medium's rules can themselves be the artwork. \emph{Hate/Move} ultimately stands as a conceptual punchline with a serious core: an artwork about authority, delivered in the medium of authority itself.

\paragraph{Key Statistics} In November 2021, Pak airdropped 30 one-of-a-kind \textit{Hate} NFTs (depicted as a simple heart symbol) to selected critics. Each had an embedded ``move'' mechanism allowing Pak to transfer them. The contract (address \texttt{0xMove…}) was written by Manifold Studio and made public on Etherscan. During the first week, at least 2 instances of Pak using the move function were observed, and 0 transfers by holders (since none could occur). The stunt quickly spread awareness; tweets about ``Pak's Hate'' garnered thousands of impressions, and articles in crypto media (e.g., CryptoTimes and others) explained the phenomenon. The term ``Invisible Mechanism'' was coined by Pak in a tweet announcing the project's true purpose. Ultimately, no monetary exchange took place for these NFTs on the market (until perhaps after Pak later unlocked them, if Pak did). Instead, their value was purely conceptual. Pak's ``Move'' contract was later referenced in discussions on NFT standards, making these 30 tokens and their story a small but significant footnote in NFT history about creative smart contract design.

\paragraph{Annotation} \emph{Invisible Mechanisms} is an outlier that inverts nearly every pattern in the portfolio. Where other works distribute control (\textbf{C2}), this one concentrates it absolutely: the \texttt{move()} function grants the artist unilateral power to seize or relocate any token. Where other works invite willing participation (\textbf{C3}), here the recipients were enlisted without full knowledge of the token's constraints. The work's true innovation, however, lies in its creation of a non-transferable asset---a token that, once deposited in a wallet, cannot be moved by its holder. This is a radical contribution to \textbf{C1} (system-centric composition): the contract's permissions structure \emph{is} the artistic statement, and the meaning of the work resides entirely in what the protocol forbids rather than what it enables. The non-transferability mechanism directly anticipates the ``soulbound token'' concept later formalized by Weyl, Ohlhaver, and Buterin \cite{Ohlhaver2022Decentralized}, making \emph{Hate} a historically significant precedent in which an artistic experiment prefigured a major theoretical development in blockchain design. The work's primary contribution is to \textbf{C6} (poetic message embedding): the non-transferable ``Hate'' token literalizes the idea that negative attention, once given, cannot be taken back---a pointed commentary on power asymmetry encoded directly in the smart contract. Furthermore, the mechanism pioneered here did not remain isolated: the non-transferability logic later inspired the soul-bound design deployed in \emph{Censored}, where message tokens were locked until Assange's release, demonstrating that even Pak's most provocative experiments serve as compositional building blocks for subsequent works.

\paragraph{Information}

\begin{itemize}
\item 
Launch Website: Pak's Twitter
\item 
Release Date: Nov, 2021
\item 
OpenSea: \url{https://opensea.io/collection/coic}
\item 
Contract Address: 
\\0x938e95271311641dc88fedaa6d7b9afdc875daa9 (ERC721)
\end{itemize}

\subsection{Merge}

\begin{figure}
    \centering
    \includegraphics[width=0.9\linewidth]{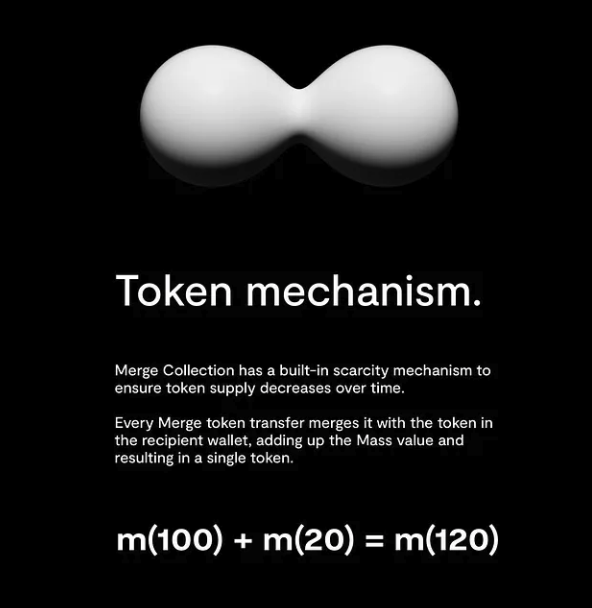}
    \Description{A screenshot from the Merge launch website. At top, a 3D rendering of two white blob-like spheres pressing together, illustrating the merging concept. Below, the heading ``Token mechanism'' followed by explanatory text: ``Merge Collection has a built-in scarcity mechanism to ensure token supply decreases over time. Every Merge token transfer merges it with the token in the recipient wallet, adding up the Mass value and resulting in a single token.'' At the bottom, the formula m(100) plus m(20) equals m(120) is displayed.}
    \caption{The token mechanism explained on the \textit{Merge} (2021) launch website. When two tokens enter the same wallet, the smart contract automatically combines them into one with summed mass---a built-in deflationary protocol that ensures the total supply can only decrease over time.}
    \label{fig:merge}
\end{figure}

\begin{figure*}
    \centering
    \includegraphics[width=0.9\linewidth]{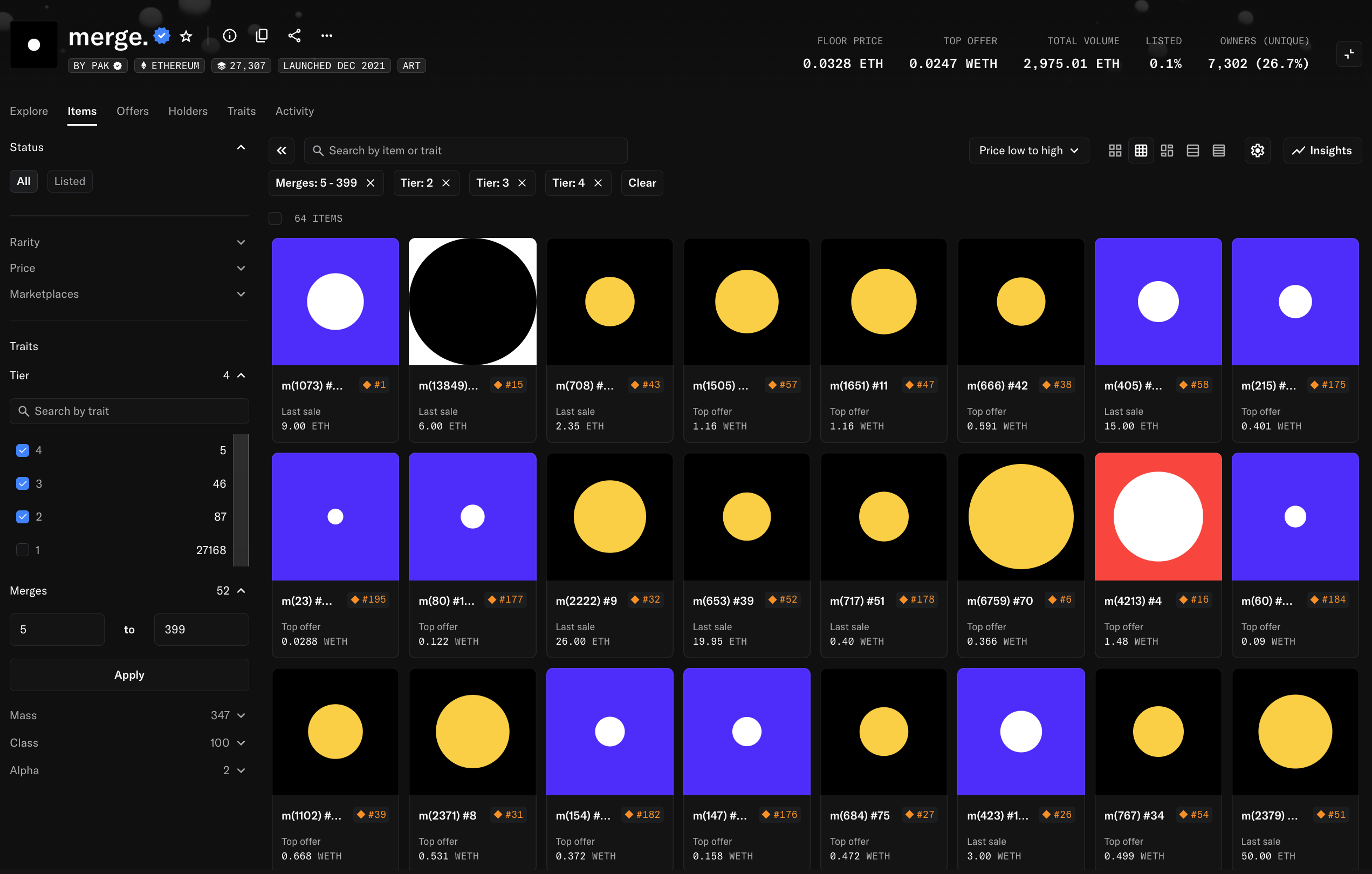}
    \Description{A screenshot of the Merge collection on OpenSea showing approximately twenty NFTs displayed as colored circles of varying sizes on colored backgrounds. Circles range from small to very large, colored white, yellow, red, and blue against blue, purple, or black backgrounds. The header shows collection stats: 27,307 items, floor price 0.0326 ETH, total volume 2,975 ETH. A sidebar displays filtering options for Tier (1 through 4), Merges count, Mass, Class, and Alpha traits.}
    \caption{\textit{Merge} on OpenSea. Each NFT is visualized as a circle whose size corresponds to its accumulated mass. Color tiers (white, yellow, red, blue) indicate mass thresholds, making the relative ``weight'' of each collector's holdings immediately legible across the marketplace.}
    \label{fig:merge-opensea}
\end{figure*}

\begin{figure*}
    \centering
    \includegraphics[width=0.9\linewidth]{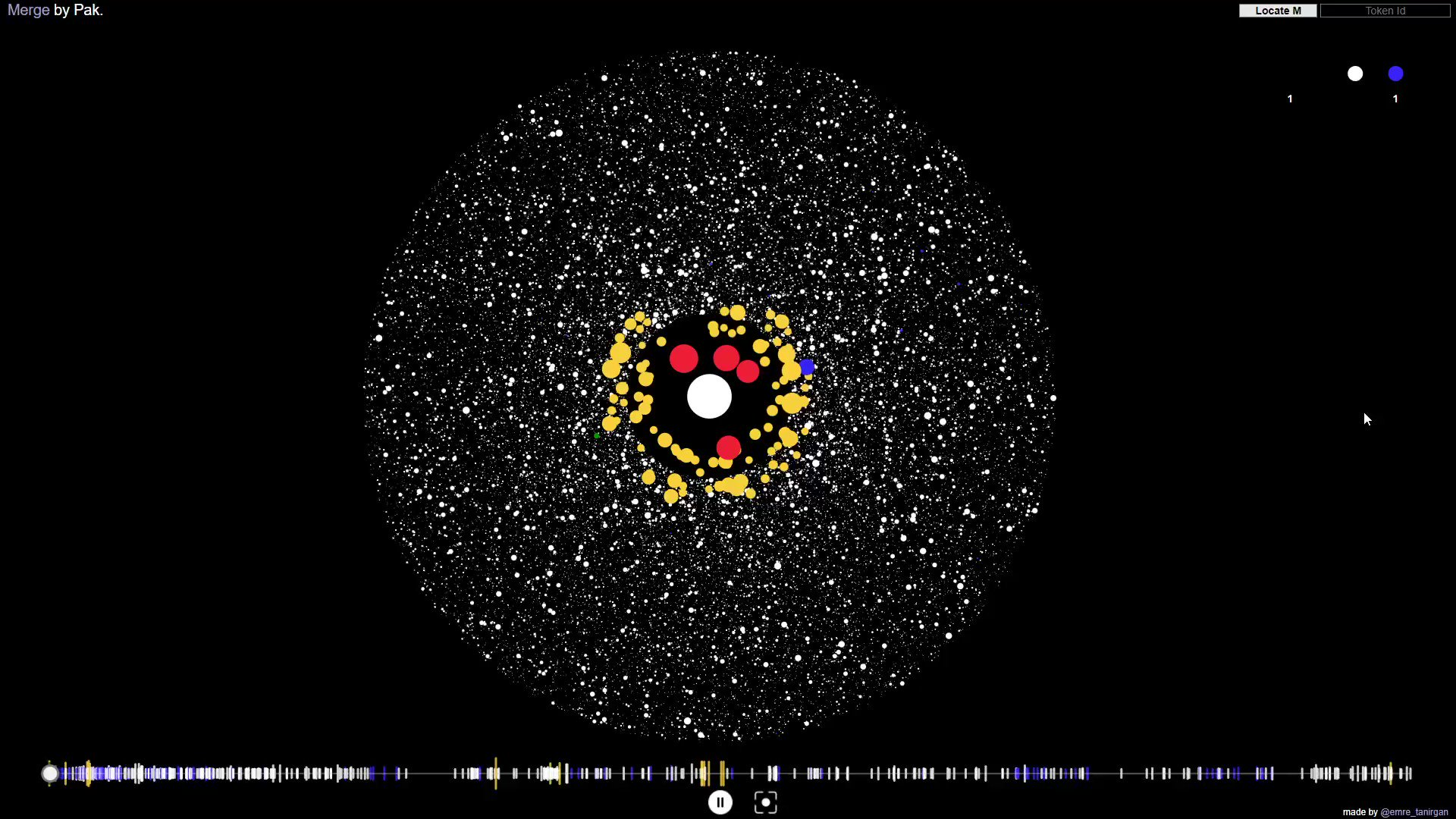}
    \Description{A third-party data visualization resembling a galaxy or solar system. Thousands of small white dots representing individual Merge tokens are arranged in a radial pattern around a cluster of larger colored spheres (white, red, yellow, blue) at the center representing the highest-mass tokens. A timeline bar along the bottom shows temporal data. The overall effect evokes a cosmic distribution of mass.}
    \caption{Third-party interactive visualization of all \textit{Merge} tokens at \url{https://mass.show/}. Each dot represents one token, sized proportionally to its mass. The galaxy-like distribution reveals the power-law concentration of mass among a few dominant holders surrounded by thousands of smaller participants.}
    \label{fig:merge_viz}
\end{figure*}

\paragraph{Concept} \emph{Merge} (December 2021) is a landmark NFT work wherein Pak explored themes of unity, rarity, and collective participation by creating an artwork that exists in fragments yet has the potential to become a single whole. Billed as the first artwork that collectors could collectively assemble, \emph{Merge} was sold not as discrete editions but as ``mass units''---small identical tokens of mass that buyers could acquire in any amount during a 48-hour sale. Each buyer's mass units automatically fused into one NFT (a single mass) in their wallet, whose size (visualized by a growing circular form) was proportional to the amount of mass purchased. The concept plays on the idea of merging: if two \emph{Merge} NFTs ever end up in the same wallet, they combine into one larger mass, reducing the total number of tokens in circulation. In effect, Pak conceived \emph{Merge} as a dynamic, self-collapsing collection---one that challenged the traditional notion of an edition size by making the supply theoretically shrinkable over time. Thematically, \emph{Merge} invokes a sense of digital unity: all collectors hold a piece of a conceptually singular artwork, and if one entity were to gather every piece, the title's promise would be realized as a single merged entity. Through this, Pak commented on the interplay between collaboration and competition, inviting the question of whether collectors would consolidate or guard their individual pieces, and what ultimate form the artwork might take.

\paragraph{System Mechanism} The \emph{Merge} contract innovated on the token model by introducing an additive property to NFTs. Technically, each mass unit was an ERC-1155 token minted during the sale; at the close of the sale, the contract ``compressed'' a buyer's multiple units into a single ERC-721 NFT that recorded the total mass count as an attribute. The NFT's visual appearance (a circle of a certain size and perhaps color) was programmatically determined by the mass count, making the artwork generative and data-driven. The merging logic was enforced at the contract level: it utilized a transfer hook that, upon detecting a \emph{Merge} token arriving in a wallet that already held one, would cancel the separate existence of the incoming token and increment the mass of the token already in that wallet. This mechanism guaranteed that no wallet could ever hold more than one \emph{Merge} NFT---a radical departure from standard NFTs. Over time, if collectors consolidated holdings or accidentally merged by buying with a pre-existing mass in their wallet, the total token count of the project would decrease from the initial 28,983 supply. Unlike most art, where the edition size is fixed, \emph{Merge} had a fluid supply that could contract, theoretically even to 1. This dynamic introduced emergent phenomena: for example, the rarity of certain visual variants depended on how people managed their tokens, not solely on predetermined traits. By embedding these rules, Pak effectively encoded a set of interactions and potential outcomes (even conflict) into the artwork's DNA, making \emph{Merge} as much a social experiment as a digital sculpture.

\paragraph{Participatory Interaction} Participation in \emph{Merge} was straightforward yet unprecedented in scale: 28,983 collectors took part in the open sale, collectively purchasing 312,686 mass units over the two-day period. Instead of competing for limited editions, buyers were cooperating in a sense---everyone was guaranteed to receive their own NFT mass, and many aimed to accumulate as much mass as they could. The interactive twist emerged after the sale: whenever a \emph{Merge} NFT was transferred on the secondary market, the contract would check the recipient's wallet. If the recipient already had a \emph{Merge} NFT, the incoming mass would merge with the existing one, destroying one token and increasing the other's mass. This meant that collectors had to strategize; some chose to split their holdings across multiple wallets to keep separate smaller masses (especially if those had unique visual traits like a different color), while others intentionally merged to build a singular, more ``massive'' mass. There was even a gamified angle of aggression and defense: theoretically, someone could send a large mass to another collector's address (without permission) to forcibly merge and absorb a smaller mass, as a way to sabotage its rarity or uniqueness. Such scenarios made holding a \emph{Merge} NFT a participatory experience beyond the initial purchase: the community was actively discussing merger tactics, bragging about the size of their mass, and contemplating alliances or friendly competitions. In essence, \emph{Merge} turned collecting into a collective game that extended into every transaction post-drop.

\paragraph{Artist's Control} After \emph{Merge} was launched, Pak's role was largely hands-off; the rules in the smart contract dictated the fate of the tokens. Pak did not exert control over who could buy or how much (beyond setting the sale's time frame and pricing increments), nor could Pak intervene in the merging process once tokens were in circulation---that process was entirely automatic and irrevocable. However, Pak's control was asserted in the careful design: Pak set the initial conditions that governed everything, from the pricing mechanism (which increased the price per mass unit in stages as certain sales milestones were hit) to the visual algorithm that represented mass. By relinquishing active control and entrusting the artwork to self-execution, Pak reinforced \emph{Merge}'s stance as a decentralized piece governed by protocol. In fact, the only way Pak influenced the outcome was indirectly, through Pak's initial decision not to cap the edition: this allowed the community's demand to define the scale of the piece (and ultimately break the record for the largest-ever art sale by a living artist). In the life of \emph{Merge} post-sale, Pak stepped back completely---any further evolution (like tokens merging into fewer tokens) was purely at the discretion of the holders and the mechanics Pak had put in place. This minimal ongoing control is consistent with Pak's tendency to let Pak's protocol-based works run their course autonomously, in stark contrast to the interventionist stance of \emph{Hate/Move}.

\paragraph{Collective Emergent Behavior} The launch of \emph{Merge} saw an enormous turnout, and this critical mass of participants created a unique collector community bonded by the experiment. Immediately, \emph{Merge} holders began comparing the sizes of their masses, fostering playful rivalries between ``whales'' (those with very large masses) and everyday collectors. There was a leaderboard mentality; some individuals who acquired huge quantities of mass became minor celebrities in the community, their holdings seen as a bold statement of support for Pak's vision. On secondary markets, novel behaviors emerged: because any purchase that combined two masses was irreversible, some collectors were hesitant to buy an additional \emph{Merge} token unless it was larger than the one they already possessed, to avoid `losing' their smaller piece in a merge. This led to an unusual trading strategy where sometimes whole wallets (with a \emph{Merge} NFT inside) were sold peer-to-peer, just to preserve the token's separate identity. The theoretical endgame of one entity eventually merging all masses into ``The One'' became a topic of both amusement and intrigue, serving as a metaphor for consolidation of power or unity of art. Meanwhile, the sheer fact that \emph{Merge} had broken the world record for an art sale---not by one person's expenditure but by the collective effort of nearly 29,000 buyers---instilled a sense of communal pride. This remains a profound aspect of the piece's narrative: it proved that a decentralized network of collectors could, together, make art history. Over time, as some merges did occur and the token count slowly diminished, the community continued to monitor and discuss these developments, treating the artwork as an ongoing story they were all part of. \emph{Merge} thus nurtured both camaraderie and competition among its participants, encapsulating a microcosm of the crypto community's spirit.

\paragraph{Poetic Meaning-Making} On a conceptual level, \emph{Merge} operates as a poetic exploration of unity and the power of the many. It turned the act of collecting into a narrative of convergence: each individual NFT was not just an endpoint but a piece of a larger potential composition. There is an inherent poetic image in thousands of separate collectors holding what could be fragments of one singular artwork---it's as if a digital star was shattered and distributed, with the ever-present possibility of reassembly. This resonates with ideas of community in the crypto space: \emph{Merge} implicitly asks whether the true value of an artwork might reside in shared ownership and collective action rather than exclusivity. The artwork's very existence and record-breaking price were a testament to collaboration (a crowd of buyers rather than a single bidder), making it a celebration of a decentralized patronage model. Additionally, the merging mechanic itself can be seen as a commentary on accumulation and synthesis: smaller masses being absorbed by bigger ones evokes phenomena in economics and nature alike, delivered here as a voluntary game. By provoking participants to consider merging or resisting merges, Pak surfaced questions about competition versus cooperation and about whether art is better experienced as a multitude or as a unity. Ultimately, \emph{Merge}'s broader impact lies in its demonstration that scarcity and unity can be artistically intertwined: it delivered a visual and participatory poem about how the many can become one. In doing so, it expanded the discourse of digital art to include not just the creation of images, but the creation of new social contracts and economic structures as vehicles for meaning.

\paragraph{Key Statistics} \textit{Merge} was sold on 2--4 December 2021 and drew 28,983 buyers who purchased 312,686 total mass units. The sale grossed \$91.8 million, making it (at that time) the highest total for any NFT artwork and thrusting Pak above Jeff Koons as the priciest living artist by primary market sales. The average collector bought about 10.8 mass units. Initially, 28,983 Merge tokens existed (one per buyer). Through secondary-market activity and merging, the supply has been decreasing: one year post-sale, the number of distinct Merge tokens had dropped to around 27,000 as consolidations happened. The largest token (\textit{Alpha}) amassed 933,878.2 mass (some collectors found ways to add mass beyond the sale via special mechanics or bonuses), and the next few largest were orders of magnitude smaller, highlighting a steep consolidation curve. The smart contract ensured that a Merge token's mass count and visual size update in real-time with each merge; Nifty Gateway's interface had to be adapted to handle these dynamic NFTs. By January 2022, \textit{Merge} tokens had done over \$100 million in secondary trading volume as collectors continued to jockey for position. Importantly, no single entity has (yet) merged all tokens---tens of thousands of decentralized pieces remain, meaning \textit{Merge} lives on as a plural artwork owned by many, with the theoretical possibility that it could one day coalesce further \cite{Thomas2021Pak}.

\paragraph{Annotation} \emph{Merge} is the most ambitious and conceptually far-reaching work in the portfolio, serving as the definitive case for \textbf{C1} (system-centric composition) and \textbf{C2} (autonomous governance): once deployed, its self-merging deflationary logic operates without any artist intervention, and the artwork's visual form (the size and color of each mass token) is entirely determined by on-chain state. After the initial open-edition sale, everything becomes an automatically self-governed universe in which the protocol alone dictates outcomes. \textbf{C3} (distributed agency) reaches its largest scale here---28,983 buyers collectively sculpt the work's final distribution---while \textbf{C5} (economy-driven engagement) manifests not only in the frenzied mass-hoarding competition but in the emergence of the Internship DAO, a collectively organized effort to acquire the Alpha token that arose entirely from participant initiative rather than artist design. This emergent behavior is paradigmatic: user activity creates its own attentional dynamics, as though the work generates a distinct game that no one scripted. Every person, at every moment, faces the choice to sell or to acquire others' tokens, and this perpetual decision space---playable at any time, but always with genuine economic incentive---transforms \emph{Merge} into something closer to a persistent world than a static artwork. Merge is notable for the \emph{irreversibility} of its dynamics: tokens can only grow through merging, never split, creating a one-directional ``extinction game'' that distinguishes it from the reversible or cyclical mechanics of burn.art and Lost Poets. This irreversibility gives Merge a distinctive temporal character (\textbf{C4}): the work can only become more concentrated over time, never return to its original state, ensuring that its aesthetic and economic trajectory is entropic and unrepeatable.

\paragraph{Information}

\begin{itemize}
\item 
Launch Website: \url{https://www.niftygateway.com/collections/pakmerge/}
\item 
Release Date: Dec 2, 2021
\item 
OpenSea: \url{https://opensea.io/collection/m}
\item 
Contract Address: 
\\0xc3f8a0f5841abff777d3eefa5047e8d413a1c9ab (ERC721)
\end{itemize}

\subsection{Censored}
\begin{figure}[ht]
    \centering
    \begin{subfigure}{0.8\linewidth}
    \centering
    \includegraphics[width=0.9\linewidth]{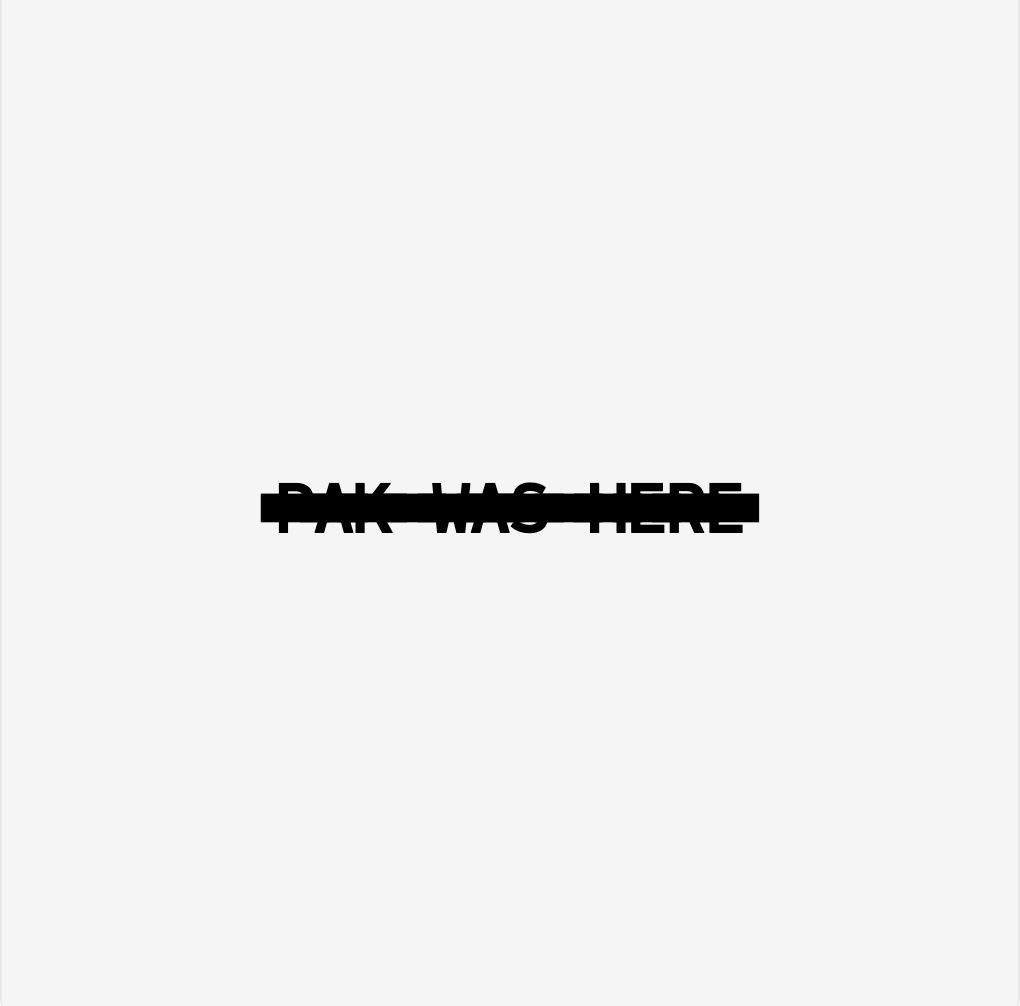}
    \Description{A light gray background with the text ``PAK WAS HERE'' partially visible behind a thick black horizontal redaction bar that crosses through all three words, obscuring but not fully hiding the message.}
      \end{subfigure}
  \begin{subfigure}{0.8\linewidth}
    \centering
      \vspace{10pt}
    \includegraphics[width=0.9\linewidth]{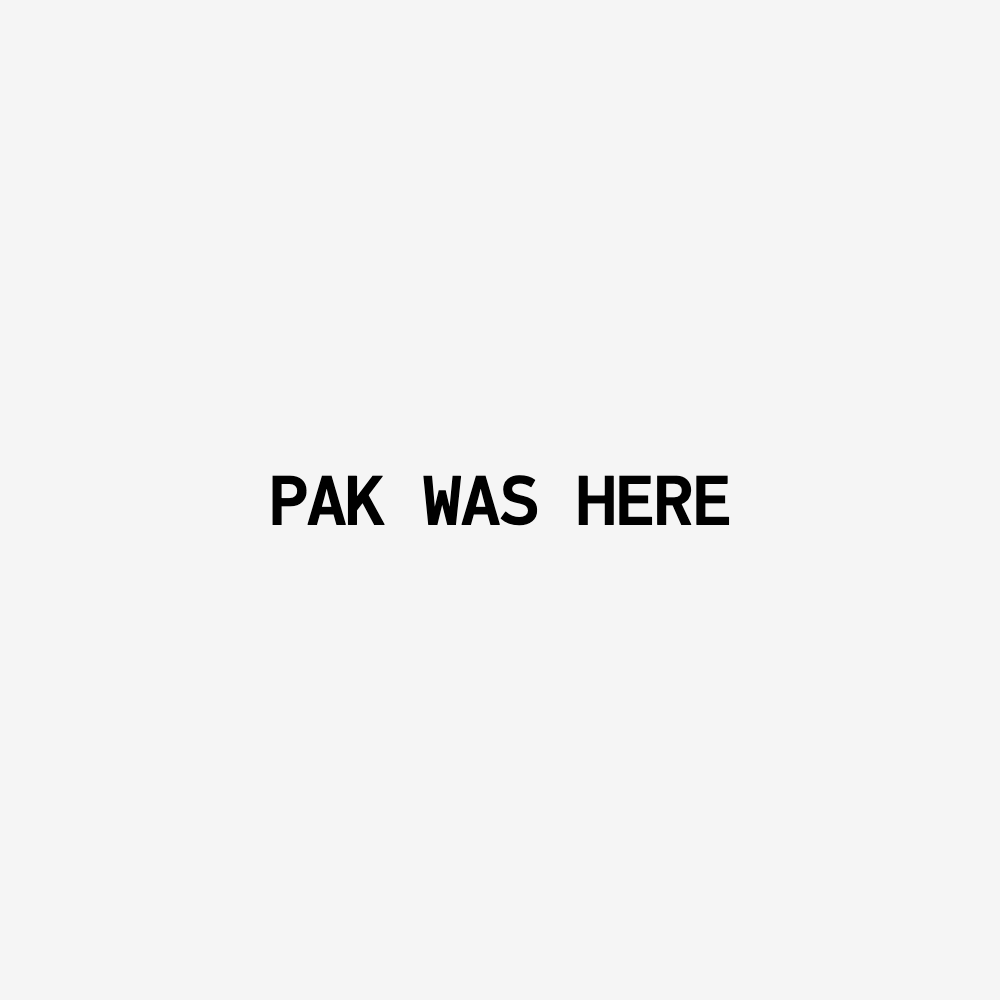}
    \Description{A light gray background with the text ``PAK WAS HERE'' displayed clearly in bold black sans-serif capital letters, with no redaction bar, showing the fully revealed uncensored state of the NFT.}
  \end{subfigure}

    \caption{``Censored is a Collection by Pak and Assange and You.'' \textit{Censored} (2022). Top: a soul-bound message NFT in its censored state, with text obscured by a redaction bar---untransferable while Assange remains imprisoned. Bottom: the same NFT in its uncensored state after Assange's release, revealing the full message and becoming freely transferable.}
    \label{fig:censored}
\end{figure}

\paragraph{Concept} \emph{Censored} (February 2022) is a two-part collaborative artwork by Pak and WikiLeaks founder Julian Assange that confronts issues of censorship, free speech, and the power of the collective. Conceived while Assange was imprisoned, the project's central question is how information and truth can be either suppressed or liberated in the digital age. The first part, \emph{Clock}, is a one-of-a-kind dynamic NFT that simply displays a count of the number of days Assange has been in custody---an ever-increasing tally that starkly visualizes the passage of time lost to censorship and incarceration \cite{Censored2022}. The second part was a dynamic open edition often referred to as \emph{Censored} (later ``Uncensored''), which invited anyone to anonymously submit a text message. Each submission was minted as an NFT where the message initially appeared as blacked-out blocks (as if redacted), but later these NFTs were revealed to show the full message---creating a permanent, public ledger of uncensored speech on the blockchain. Together, these components embody a powerful concept: \emph{Censored} is both a protest and a platform, highlighting the plight of Assange specifically and the principle of uncensored expression universally. By design, the project blurred the line between artist and participant, making ``you''---the public---a co-creator of the open edition piece, echoing the idea that free expression is a collective, democratic effort.

\paragraph{System Mechanism} The technical structure of \emph{Censored} mirrored its conceptual duality. \emph{Clock} was implemented as a dynamic NFT: its content updates once per day to increment the count of Assange's imprisonment. This likely involves either an on-chain calculation of days since a start date or an off-chain oracle pushing daily updates---either way, the piece is programmatically tied to the real-world passage of time and will continue to tick until Assange's status changes. The open edition was dynamic in a different sense: when each \emph{Censored} NFT was first minted, the text submitted by the user was stored (on-chain or via metadata) but visually represented in a redacted style (for example, as black bars or scrambled characters). Then, upon Assange's release on 24 June 2024, a contract function executed to ``uncensor'' all these tokens, updating the metadata so that the true text became visible. Each open-edition NFT is thus unique in message but identical in having undergone the same transformation from obscured to revealed. The fundraising mechanism was straightforward: the sale smart contract directed the incoming Ether to designated wallets associated with Assange's defense fund and other aligned charities (indeed, a portion of the 671 ETH raised from the open edition was later donated to support journalistic freedom and even humanitarian aid). The dynamics here combined artistic intent with real-world impact: the more people voiced themselves (minted NFTs) and the higher the DAO bid for \emph{Clock}, the more tangible support was generated for the cause. In a sense, the blockchain functions as both canvas and conduit in \emph{Censored}, ensuring that the art's message (the imperative of free expression) is inseparable from the action it precipitates (material aid for a free press).
\paragraph{Participatory Interaction} The participatory element of \emph{Censored} was striking. For \emph{Clock}, participation took the form of collective fundraising: an Assange-supporter DAO (decentralized autonomous organization) pooled resources from over 10,000 people to bid on and ultimately win the piece for 16,593 ETH (about \$52.8 million), making each contributor a part-owner of the artwork and massively amplifying its political message. For the open edition, Pak opened a web portal on February 7, 2022, where anyone could type a short message and mint it as an NFT. There were minimal barriers---aside from blockchain transaction fees---which allowed thousands of ordinary people worldwide to inscribe their uncensored thoughts onto the blockchain. During the mint period, participants wrote messages ranging from political statements (``FREE ASSANGE NOW'') to personal reflections, fully aware that they would initially be recorded in a censored (illegible) state. After the mint closed, Pak triggered a global reveal: all the NFTs' images were updated to show the actual text of the messages, symbolizing voices breaking through suppression. In this way, the audience's role wasn't just to observe but to speak and be memorialized as part of the art. Additionally, by buying these open-edition NFTs (each priced modestly), participants contributed to a fundraiser---all proceeds were directed to Assange's legal defense and pro-freedom organizations---blending activism with artistic engagement. In essence, \emph{Censored} turned its audience into an active assembly of publishers and protesters, demonstrating the principle that in a decentralized network, everyone can raise their voice and collectively bolster a cause.

\paragraph{Artist's Control} In \emph{Censored}, Pak acted not only as an artist but also as a facilitator of collective expression and activism. Pak and Assange's team set the stage---defining the parameters of the open edition (time frame, interface for input, price) and crafting the \emph{Clock} NFT---but then handed the mike to the public. For the open edition, Pak exercised restraint in content moderation: applying no filters in alignment with the project's absolutist free speech stance, Pak allowed minters to submit any text, which was a radical trust in the community. However, Pak retained control over the overall execution: Pak decided when the reveal would happen and ensured it occurred uniformly for everyone, and Pak controlled the smart contract that forwarded funds to the intended beneficiaries. With \emph{Clock}, once it was launched, the control shifted to the AssangeDAO (as the new owners) in terms of display and eventual fate of the piece, although the daily increment feature was baked into the token itself. In essence, Pak's control was about creating a secure and open channel for expression and then stepping back at the right moments. The success of \emph{Censored} depended on this light-touch approach; too much control (e.g., censoring the messages or manipulating the outcomes) would have undermined the trust and authenticity the piece needed to resonate. By structuring it as Pak did, Pak underscored Pak's role as an enabler---Pak controlled the framework but not the voices within, embodying the artwork's ethos in the very way it was managed.

\paragraph{Collective Emergent Behavior} The \emph{Censored} project galvanized a uniquely blended community of art enthusiasts, cypherpunks, transparency advocates, and supporters of Assange. The formation of AssangeDAO itself was a landmark event: thousands of strangers coordinated in a matter of days on chat platforms, raised tens of millions in cryptocurrency, and collectively won the \emph{Clock} auction, an unprecedented feat of decentralized crowd-funding. This feat gave participants a profound sense of accomplishment---each member of the DAO owns a fractional piece of \emph{Clock} via tokens, making \emph{Clock} a collectively owned symbol of protest. In parallel, the open edition saw 29,766 NFTs minted across its 48-hour run, reflecting broad participation from the crypto community and beyond. Upon the reveal, a new collective experience began: people scoured the now-uncensored messages, sharing poignant or powerful quotes they found among the NFTs. The collection became a digital chorus of statements about freedom, justice, and personal sentiments, and owning one meant being part of that chorus. A notable emergent behavior was the way traditional lines blurred: activists became art collectors by minting NFTs, and NFT collectors became activists by joining the political messaging. The \emph{Censored} Discord and other forums buzzed with discussion not just about the art's value, but about the cause it supported and the stories behind individual messages. This convergence of normally separate communities (art, crypto, activism) was itself a product of the project. In the aftermath, the collective energy didn't dissipate immediately: the DAO refocused on continued advocacy for Assange, and many open-edition holders continued to promote the messages they had minted, effectively using their NFTs as badges of alignment with the free speech movement. \emph{Censored} thus forged an alliance of convenience into a lasting community centered on principle and expression.

\paragraph{Poetic Meaning-Making} \emph{Censored} stands out as a poignant union of art and activism, transforming the cold mechanics of blockchain into an emotional and political narrative. The poetic substance of the work arises from its contrasts: a single, ever-ticking clock counting the days of one man's silencing, and a sea of uncensorable messages giving voice to thousands. The \emph{Clock} NFT is minimalistic yet profound---each passing day it displays is a testimony to injustice, a visual poem of waiting and resilience. The open edition, once uncensored, became an anthology of global voices; its poetry is both literal (many entries were written with genuine passion or wit) and metaphorical, in that it turned censorship on its head---what was once hidden is now immortalized in public view. There is a strong sentiment that \emph{Censored} wasn't just documenting a moment in crypto art, but a moment in history: it bridged the gap between the decentralized art world and real-world social issues, demonstrating that the blockchain community could rally around something deeply human. The broader impact of \emph{Censored} lies in its demonstration of how digital art can be immediately socially relevant. It set a precedent for using NFTs as a medium of protest and fundraising, expanding the notion of what kind of statements can be made---and preserved---through art. In summary, \emph{Censored} made poetic meaning through direct action: it is art as a verb, an event in which aesthetic expression, technological infrastructure, and moral conviction converged to powerful effect.

\paragraph{Key Statistics} The \textit{Clock} NFT sold on 9 February 2022 for 16,593 ETH (approximately \$52.8 million), bought by AssangeDAO, a collective of 10,000+ members pooling funds. This made it one of the most expensive NFTs ever and provided a huge donation to Assange's legal defense. The open edition \textit{X/X} minted 29,766 NFTs over its 48-hour run, raising 671 ETH in voluntary contributions (about \$2.1 million) for pro-freedom organizations chosen by Pak and Assange. Each of these message NFTs was initially non-transferable; following Assange's release on 24 June 2024, the tokens were unlocked and the censored messages revealed. Over 17,000 unique Ethereum addresses participated in the mint (many contributed multiple messages via additional wallets). The largest donation from a single minter was 50 ETH (showing some used the mint as a way to donate significantly). Post-drop, AssangeDAO has kept \textit{Clock} on display via a fractionalized ownership governance token (representing membership shares in the DAO). \emph{Censored} stands as a record-breaking instance of political fundraising through digital art, and a case where an NFT project directly engaged international press and communities far beyond the traditional art world.

\paragraph{Annotation} \emph{Censored} is the most politically charged and arguably the most poetically powerful work in the portfolio, serving as the strongest case for \textbf{C6} (poetic message embedding): the entire interaction ritual---submitting a message that is initially blacked out, then revealed only upon Assange's release---enacts the experience of censorship and liberation in a way that purely visual art cannot. The political claim embedded in the work is inseparable from the collective behavior it demands: nearly 30,000 participants became co-authors of a collective ``censored chorus,'' each contributing a personal message that functioned simultaneously as artistic expression and political advocacy for Assange's release (\textbf{C3}, distributed agency). The poetic message was literally embedded into the NFT, and the entire arc---from locked, non-transferable tokens bearing blacked-out text to their eventual unlocking and revelation upon Assange's release on 24 June 2024---constitutes a profoundly poetic process with genuine geopolitical meaning. The non-transferability mechanism, inherited from the soul-bound logic pioneered in \emph{Hate / Invisible Mechanisms}, here acquires a new resonance: the tokens were imprisoned alongside Assange, and their liberation mirrored his. \textbf{C2} (autonomous governance) takes an unusual form: AssangeDAO's collective ownership of Clock, and the reveal mechanism's binding to an external real-world event, make the artwork's governance partly extra-protocol---dependent on geopolitical outcomes beyond any participant's control. This external dependency is unique in the portfolio and raises the question of whether protocol art's autonomy can meaningfully extend to conditions outside the blockchain. The whole process---collective advocacy encoded in code, censorship enacted and then reversed, tokens locked and then freed---is deeply powerful, demonstrating that protocol art can achieve a form of political poetry in which the medium's constraints and affordances carry moral weight.

\paragraph{Information}

\begin{itemize}
\item 
Launch Website: \url{https://censored.xyz/}
\item 
Release Date: Feb 7, 2022
\item 
OpenSea: \url{https://opensea.io/collection/censored-pak-assange}
\item 
Contract Address: 
\\0xda22422592ee3623c8d3c40fe0059cdecf30ca79 (Soul-bound ERC721)
\end{itemize}

\subsection{Not Found / \#404 (2023)}

\begin{figure}
    \centering
    \includegraphics[width=0.9\linewidth]{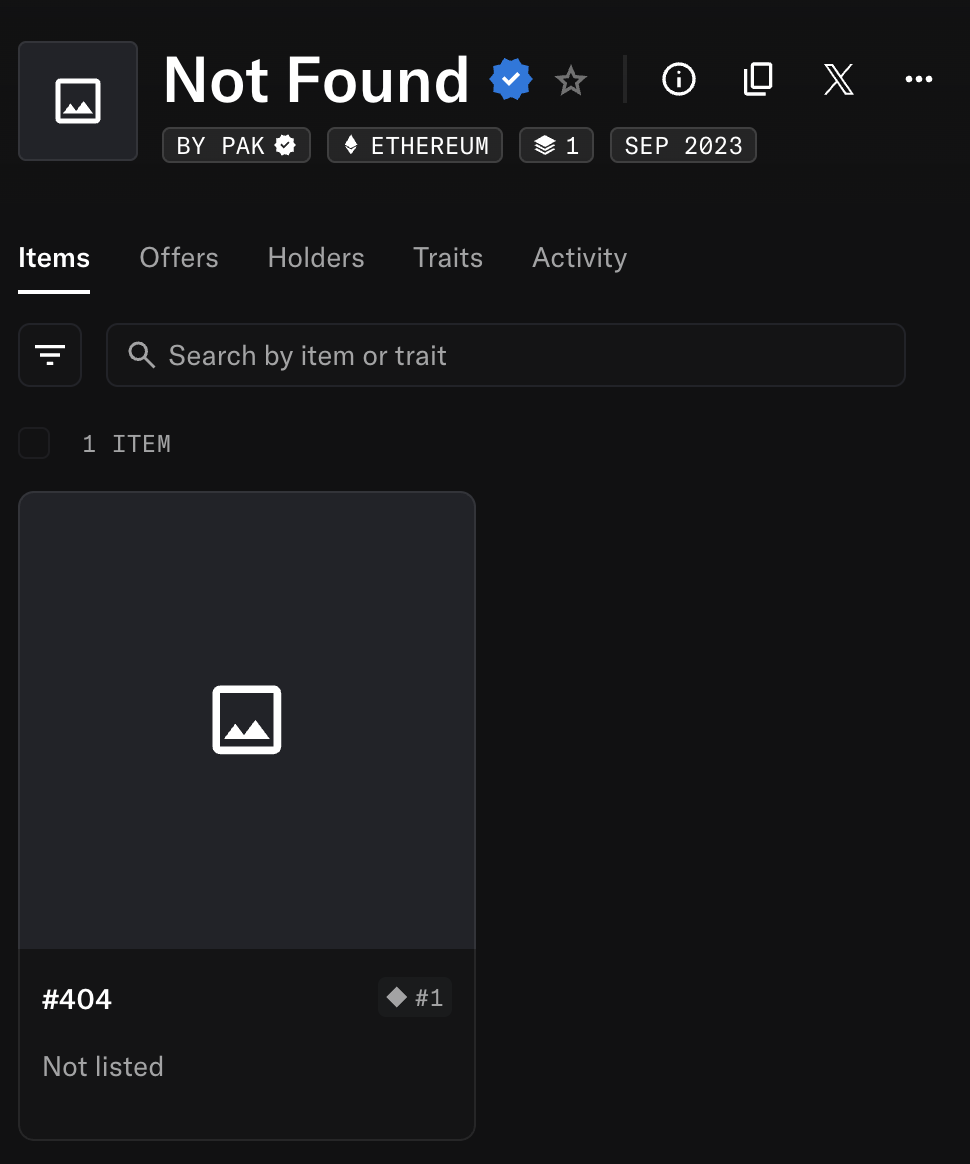}
    \Description{A screenshot of the OpenSea collection page for ``Not Found'' by Pak. The header shows the collection name with a verified badge, labeled ``BY PAK,'' ``ETHEREUM,'' ``1 item,'' and ``SEP 2023.'' Below, the single NFT entry displays only a generic broken-image placeholder icon (a small mountain landscape silhouette inside a square) on a dark background, labeled ``number 404, rank 1, Not listed.'' No artwork image, title, or description is shown.}
    \caption{\textit{Not Found} \#404 on OpenSea (2023). The token was minted with no metadata by design---no image, no name, no description---so that any platform querying it receives only a 404 error. The resulting broken-image placeholder becomes the artwork itself: a deliberate void memorializing the absent.}
    \label{fig:notfound}
\end{figure}
\paragraph{Concept} \emph{Not Found \#404} is a 1-of-1 NFT forged ``in memory of the absent'' that Pak created in August 2023 as a poignant tribute to the late crypto artist Alotta Money. The concept revolves around absence as presence. It leverages the web's familiar ``404 Not Found'' error---the message one sees when a webpage is missing---as an artistic statement. Pak minted \textit{404} with no metadata: no image, no name (beyond the token ID), no description. In Pak's words, ``\#404 is a token forged in memory of the absent\ldots{} Void of metadata by design to amplify the presence of absence.''   The NFT exists as a sort of intentional glitch or blank, symbolizing the void left by Alotta Money's passing. Socially, it speaks to how loss is felt in the digital age---even on a blockchain, an empty space can carry profound meaning as a memorial.

\paragraph{System Mechanism} Technically, \textit{404} is engineered to be an NFT that always produces a 404 error for its metadata endpoint. Typically, an NFT's metadata link provides JSON data including the name, description, and image URI. Pak deliberately set up \textit{404} so that when any application (like OpenSea or Etherscan) queries the token's metadata, the server responds with a ``Not Found'' status. This means the token has no visible content---no name, no image---just its token ID (\#404 on that contract). However, embedded within the smart contract is a tiny message: instead of the usual metadata, Pak hard-coded a plain string that says ``In memory of the absent'' as the response. That phrase is the only ``content'' of the NFT, acting like a eulogy delivered in code. The dynamic effect is that on any NFT marketplace, \textit{404} appears as an empty or broken entry (which is itself conspicuous). Over time, assuming the piece stays as is, it will forever display as a missing asset, effectively eternalizing the idea of absence. This is a case where the lack of data is the feature, not a bug---a radical use of the NFT format to signify zero.

\paragraph{Participatory Interaction} \textit{404} was part of a charity auction (the Alotta Money Memorial on MakersPlace), so the direct participation was in the bidding process. Collectors vied not for a flashy image but for the honor of owning this conceptual piece. The winning participant, prominent collector WhaleShark, essentially paid a large sum for an NFT that ``looks'' like a broken link. In a broader sense, the audience participation comes in the form of interpretation: everyone who views the token on marketplaces just sees an error or blank, and must mentally fill in what it represents. Interestingly, because the contract ensures the metadata always returns a 404 error, any platform displaying it will show a broken image icon or ``Not Found'' text, inviting each viewer to participate by remembering or inquiring about Alotta Money and the context. Thus, while not interactive in a traditional way, it engages the audience's awareness and curiosity. The community also participated in spreading the story---tweets and articles explained what \textit{404} meant, effectively crowdsourcing the task of giving this ``silent'' NFT a voice.

\paragraph{Artist's Control}
 Pak had meticulous control over \textit{404}'s presentation. By deciding to include no metadata and making the token unchangeable (immutable after mint), Pak ensures that neither Pak nor the owner can ever add an image or title later. This permanence is part of the tribute's integrity. The artist also controlled the context: \textit{404} was introduced with a tweet explaining its purpose and dedication, guiding the initial audience understanding. After the auction, however, Pak's control yielded to the token's behavior in the wild---any confusion or discussion arising from people seeing a blank NFT became part of the piece's life. It's worth noting that creating an NFT that defies normal display could have negative repercussions (some might think it's an error), but Pak accepted that risk to preserve the concept. In terms of curation, Pak placed \textit{404} in a charity event so that its sale (48 ETH) benefited a cause (Alotta's cancer fund or a related charity), aligning control of proceeds with the homage. Ultimately, Pak's control was about ensuring emptiness---an ironic but powerful form of control where doing ``nothing'' was a deliberate artistic act.

\paragraph{Collective Emergent Behavior} 
Once \textit{404} was out in the world, the NFT community responded collectively by imbuing it with significance. Discussion threads emerged about \emph{``the NFT with no data''} and what it meant; in this way, the community co-authored the narrative by sharing the backstory. Some artists were inspired by this approach and contemplated similar ``negative space'' works. As for the owner, WhaleShark displayed \textit{404} in virtual galleries and on social media, where the emptiness became thought-provoking. Rather than diminishing interest, the lack of an image actually drew people in---a reverse of the usual NFT hype cycle. A subtle emergent behavior is that marketplaces had to handle a token with no metadata gracefully; some updated their interface to show ``Unnamed'' or simply the token ID. This sparked minor technical conversations in developer circles about how to index such an NFT, thereby \textit{404} gently pushed the boundaries of NFT platform expectations. The most heartening emergent behavior was the collective act of remembrance: the NFT community, normally fixated on visuals and rarity, paused to remember an artist through a non-visual artifact. In that sense, \textit{404} succeeded in creating a communal moment of silence, as it were, in the bustling NFT space.
\paragraph{Poetic Meaning-Making} \textit{404} is pure poetry in digital form. Its poetry lies in what is absent: it forces the viewer to confront a void and find meaning in it. Much like John Cage's silent composition ``4'33''{}'' or Rauschenberg's erased drawing, Pak's \textit{404} finds art in nothingness, which in this context becomes a profound statement about loss. The use of the ``Not Found'' web error as the medium ties our sense of missing information to the emotion of missing a person. It's a requiem encoded as a glitch. The token number 404 itself is meaningful---in web lore, 404 symbolizes a dead link, something that was once there but no longer accessible, mirroring Alotta Money's departure. Moreover, \textit{404} leverages the permanence of blockchain to eternalize an impermanent idea (absence). Its message \emph{``In memory of the absent''} resonates universally---it's not just about one individual, but about all those we've lost (it invites anyone to project their own feelings of loss onto it). As a social/artistic message, it reminds the tech-forward NFT community of human mortality and the gap that death leaves in our networks. In its quiet way, \textit{404} perhaps also comments on the oversaturated NFT market---amid thousands of gaudy images, the most meaningful piece can be an empty one. The poetry of \textit{404} lies in its silence and subtlety, making the presence of an absence palpably felt.
\paragraph{Key Statistics} \textit{404} was minted on 31 August 2023 and auctioned the next day at the ``Alotta Money Tribute'' event on MakersPlace. It sold for 48 ETH (roughly \$78,000 at the time) to collector WhaleShark, with sale proceeds going to charity. The token resides on a custom contract; it is token ID 404 and notably the only token in that contract (no other token IDs exist, making 404 both the ID and the collection name). Its on-chain metadata call returns a 404 error code by design, and the token carries a brief on-chain text ``In memory of the absent.'' The piece garnered considerable media attention for an NFT---features in art publications highlighted it as the first NFT that intentionally ``does nothing'' visually yet carries deep meaning. As of 2025, WhaleShark has not listed \textit{404} for resale (and likely never will), aligning with its status as a memorial piece. \textit{404} stands as one of Pak's later protocol artworks, and while it's a single edition, its impact was amplified by the hundreds of artists and collectors who witnessed and shared in the commemorative moment it represents.

\paragraph{Annotation} \emph{Not Found} is the portfolio's limiting case: a work that strips protocol art to its conceptual minimum. \textbf{C1} (system-centric composition) is expressed through pure negation---the artwork's meaning resides entirely in the contract's deliberate refusal to return metadata, making the absence of content the content itself. \textbf{C6} (poetic message embedding) is correspondingly intense: the token ID 404 and the lone on-chain string ``In memory of the absent'' compress an entire elegy into the structure of a failed HTTP request. Unlike every other work in the portfolio, \emph{Not Found} involves no \textbf{C3} (distributed agency), no \textbf{C5} (economy-driven engagement beyond the charity auction), and no \textbf{C7} (composability). Its isolation is the point: it demonstrates that protocol art need not rely on mass participation or elaborate tokenomics to achieve poetic force---a single, precisely engineered absence can suffice.

\paragraph{Information}

\begin{itemize}
\item 
Launch Website: Pak's Twitter Account
\item 
Release Date: Aug, 2023
\item 
OpenSea: \url{https://opensea.io/collection/in-memory-of-the-absent}
\item 
Contract Address: 
\\0x3a91740d25587a0cd5baa27755876231559a3e60 (ERC721)
\end{itemize}
\end{document}